%% file: u4_ll_paper.tex
\documentclass[aps,prd,twocolumn,showpacs,superscriptaddress,letterpaper,asmmath,amssymb,nofootinbib,preprintnumbers]{revtex4-1}
\pdfoutput=1

\def \lum {1.04}

\usepackage{graphicx}
\usepackage{bm}
\usepackage{amssymb,amsmath} 
\usepackage{multirow}
\usepackage[tight]{subfigure}

\usepackage{atlas_general}

\RequirePackage{xspace}
\usepackage{relsize}

\def\Kbar  {\kern 0.2em\overline{\kern -0.2em K}{}\xspace}

\def\Bbar    {\kern 0.18em\overline{\kern -0.18em B}{}\xspace}

\def\Qbar    {\kern 0.08em\overline{\kern -0.08em Q}{}\xspace}

\def\invfb   {\ensuremath{\mbox{\,fb}^{-1}}\xspace}
\newcommand{\mev}{\ensuremath{\mathrm{\,Me\kern -0.1em V}}\xspace}
\newcommand{\mevc}{\ensuremath{{\mathrm{\,Me\kern -0.1em V\!/}c}}\xspace}
\newcommand{\mevcc}{\ensuremath{{\mathrm{\,Me\kern -0.1em V\!/}c^2}}\xspace}

\newcommand{\ttbar}{$t\bar{t}$}
\newcommand{\HT}{$H_{\rm T}$}

\def\MET{\mbox{\ensuremath{E_{\mathrm{T}}^{\mathrm{miss}}}}}

\begin{document}

\preprint{CERN-PH-EP-2012-008}
\preprint{Submitted to Physical Review D}

\title{Search for pair-produced heavy quarks decaying to $Wq$ in the two-lepton channel at $\sqrt{s}$ = 7 TeV with the ATLAS detector}

\author{The ATLAS Collaboration}

\begin{abstract}
A search is presented for heavy-quark pair production ($Q\bar{Q}$) 
under the decay hypothesis $Q\bar{Q}\rightarrow W^+qW^-\bar{q}$ with 
$q = d$, $s$, $b$ for up-type $Q$ or $q = u$, $c$ for down-type $Q$.
The search is performed with \lum{} \invfb{} of integrated luminosity 
from $pp$ collisions at $\sqrt{s}=7$ TeV collected by the \mbox{ATLAS} detector at the CERN LHC.
Dilepton final states are selected, requiring large missing transverse momentum and at least two jets.
Mass reconstruction of heavy quark candidates is performed by assuming that the $W$
boson decay products are nearly collinear.  
The data are in agreement with Standard Model expectations; 
a heavy quark with mass less than 350 GeV is excluded at 95\% confidence level.
\end{abstract}



\maketitle

\date{\today}

The addition of one or more heavy quarks is a natural extension to
the Standard Model, capable of providing an additional source of CP violation
in $B_s$ decays and accommodating a heavy Higgs boson~\cite{fourthgen, fourthgenconstraints}.
Searches for heavy quarks with the Collider Detector at Fermilab (CDF) constrain the
mass of heavy quarks ($Q$) that decay as $Q \rightarrow Wq$,
where $q = d$, $s$, $b$ for up-type $Q$ or $q = u$, $c$ for down-type $Q$,
to be \mbox{$m_{Q} >$ 340 GeV}~\cite{cdftprime}.
More specific searches have also constrained
the mass of up-type heavy quarks ($t'$) that decay as \mbox{$t' \rightarrow Wb$}
to be \mbox{$m_{t'} >$ 358 GeV}~\cite{cdftprime}
and the mass of down-type heavy quarks ($b'$) decaying via \mbox{$b' \rightarrow Wt$} 
to be \mbox{$m_{b'}>$ 372 GeV}~\cite{cdfbprime}.
The D0 experiment at Fermilab has set a mass limit of \mbox{$m_{Q} >$ 285 GeV}~\cite{d0tprime}
on heavy quarks that decay as \mbox{$Q \rightarrow Wq$}.
All previous searches used the ``lepton+jets'' channel, where only one of the
produced $W$ bosons decays hadronically.

In this Article, a search is presented for pair production of a heavy quark ($Q\bar{Q}$)
in data corresponding to an integrated luminosity of \lum{} \invfb{}
from proton-proton collisions at \mbox{$\sqrt{s}$ = 7 TeV}
collected by the \mbox{ATLAS} experiment.
The heavy quark is assumed to decay via $Q\rightarrow Wq$
where $q = d$, $s$, $b$ for up-type $Q$ or $q = u$, $c$ for down-type $Q$.
This search does not include states with $q = t$, i.e.~$d' \rightarrow Wt$ decays are assumed not to happen.
The search is performed in the dilepton channel, where both $W$ bosons decay leptonically.
Fourth-generation up-type quarks ($t'$) decaying through weak charged currents 
(\mbox{$t' \rightarrow Wb$}, \mbox{$t' \rightarrow Ws$}, and \mbox{$t' \rightarrow Wd$}) are used as a benchmark.

Three complementary searches for fourth-generation quarks were performed with the \mbox{ATLAS}
detector using \lum{} \invfb{} of 2011 data. These searches all implement $b$-quark identification
algorithms and are thus targeted towards more specific heavy quark decay modes. The channels
considered are 
\mbox{$t'\bar{t'}\rightarrow W^+W^-b\bar{b}$} in the lepton plus jets channel (setting a limit \mbox{$m_{t'} > 404$ GeV}~\cite{NewTPrimeSearch}),
\mbox{$b'\bar{b'}\rightarrow W^+W^-t\bar{t}\rightarrow W^+W^-W^+W^-b\bar{b}$} 
in the lepton plus jets channel (\mbox{$m_{b'} > 480$ GeV}~\cite{NewBPrimeSearch}), and
\mbox{$b'\bar{b'}\rightarrow W^+W^-t\bar{t}\rightarrow W^+W^-W^+W^-b\bar{b}$} 
with two same sign leptons in the final state (\mbox{$m_{b'} > 450$ GeV}~\cite{NewBPrimeSameSignSearch})."

The dileptonic final state arises in a way similar to that of pair-produced top quarks:
\mbox{$Q\bar{Q} \rightarrow \ell^+\nu q\ell^-\bar{\nu}\bar{q}$}, where $\ell$ is either $e$ or $\mu$.
Leptonically-decaying intermediary $\tau$ leptons are able to contribute to this final state if additional neutrinos are considered.
The signature is at least two jets, two oppositely charged leptons, 
and missing transverse momentum (\MET{}) from undetected neutrinos.  
Top quark pair production is the dominant source of background. 
To distinguish a potential heavy-quark signal, heavy-quark 
mass reconstruction is performed by taking advantage of the larger boost
each $W$ boson receives from the decay of a heavy quark compared to the
decay of a top quark. 
This large boost makes each undetected neutrino
approximately collinear with an observed charged lepton.

The following sections contain descriptions of the \mbox{ATLAS} detector (Section~\ref{sec:ATLASDet}), 
simulated samples (Section~\ref{sec:Simulation}), 
object reconstruction (Section~\ref{sec:ObjEvtReco}), 
baseline event selection (Section~\ref{sec:BaselineEventSel}),
data-driven background estimates (Section~\ref{sec:DDSec}), 
mass reconstruction strategy (Section~\ref{sec:MassReco}),
validation of background modeling (Section~\ref{sec:BkgVal}), 
final event selection (Section~\ref{sec:FinalEventSel}),
binned maximum-likelihood ratio fit using the reconstructed mass (Section~\ref{sec:Systematics} and Section~\ref{sec:Results}), 
and final results (Section~\ref{sec:Conclusion}).

\section{The ATLAS detector}
\label{sec:ATLASDet}

The ATLAS detector~\cite{atlasdet} is a multi-purpose particle detector
with precision trackers, calorimeters and muon spectrometers.
The momenta of charged particles with pseudorapidity \mbox{$|\eta| <$ 2.5}\footnote{ATLAS uses a right-handed coordinate system with its origin at the nominal interaction point (IP) in the center of the detector and the $z$-axis along the beam pipe. 
The $x$-axis points from the IP to the center of the Large Hadron Collider (LHC) ring and the $y$-axis points upward. 
Cylindrical coordinates $(r,\phi)$ are used in the transverse plane, $\phi$ being the azimuthal angle around the beam pipe. 
The pseudorapidity is defined in terms of the polar angle $\theta$ as $\eta=-\ln\tan(\theta/2)$.}
are measured by the inner detector (ID), 
which is a combination of silicon pixels, silicon microstrips
and a straw-tube tracker. The ID operates in a uniform 2~T axial magnetic field produced by a superconducting solenoid.
The pixel detector measurements enable precise determination of production vertices. 

Electromagnetic (EM) calorimetry for electron and photon reconstruction 
is provided by a high-granularity, three layer liquid argon (LAr) sampling calorimeter with lead absorbers 
in the region \mbox{$|\eta| <$ 3.2}.
A presampler is used to correct for energy lost by electrons and photons in material in front of the
calorimeter for \mbox{$|\eta| <$ 1.8}.
Hadronic calorimetry for \mbox{$|\eta| <$ 1.7} is provided by a scintillating tile sampling calorimeter with steel absorbers,
and for \mbox{1.5 $< |\eta| <$ 3.2} it is provided by a LAr sampling calorimeter with copper-plate absorbers.

Muons are detected with a
multi-system muon spectrometer (MS).  Precision measurements in the $\eta$ coordinate are provided by
monitored drift tubes for \mbox{$|\eta| <$ 2.7}. These are supplemented by cathode-strip chambers
measuring both the $\eta$ and azimuth ($\phi$) coordinates for \mbox{2.0 $< |\eta| <$ 2.7}.
Fast measurements required for initiating trigger logic are provided by
resistive-plate chambers for \mbox{$|\eta| <$ 1.05} and then by thin-gap chambers for \mbox{1.05 $< |\eta| <$ 2.4}.
The muon detectors operate in a non-uniform toroidal magnetic field generated by a superconducting air-core magnet system.

The ATLAS detector uses a three-level trigger system to select events for offline analysis.  
For this search, events are required to have
at least one lepton satisfying trigger requirements.
Electron trigger candidates must have transverse energy \mbox{$E_{\rm T}>$ 20 GeV}, must
satisfy shower-shape requirements~\cite{egamma} and must have an ID track matched to the EM shower.
Muon trigger candidates must have transverse momentum \mbox{$p_{\rm T}>$ 18 GeV}
and matching tracks in the ID and MS.

\section{Simulated Signal and Background Samples}
\label{sec:Simulation}

Simulated samples are used to evaluate the contributions
from the $Q\bar{Q}$ signal (assuming an up-type heavy quark) and most background processes.  
Unless otherwise noted, all events are showered and hadronized with {\sc herwig} v6.5~\cite{herwig, herwig2},
using {\sc jimmy}~\cite{jimmy} for the underlying event model. 
After event generation, all samples are processed with the {\sc geant4}-based~\cite{GEANT4} simulation of the 
\mbox{ATLAS} detector~\cite{ATLASDETECTOR} and subject to the same reconstruction algorithms as the data.

The CERN LHC instantaneous luminosity varied during data-taking
from about \mbox{$2 \times 10^{32}$ cm$^{-2}$s$^{-1}$} to \mbox{$1 \times 10^{33}$ cm$^{-2}$s$^{-1}$}~\cite{LumiPaper, LumiUncertainty}.
At maximum luminosity numerous proton-proton ($pp$) interactions were superimposed in each bunch crossing.
This pileup background produces additional activity in
the detector, affecting variables such as jet reconstruction and
isolation energies. 
Monte Carlo (MC) events simulate the pileup background by adding minimum bias events
on top of the hard scatter. 
The MC events are later reweighted such that the simulated
instantaneous luminosity distribution matches that in data.

\subsection{Heavy-quark pair production}

Production and decay of heavy-quark pairs ($Q\bar{Q}$) is modeled with 
the leading-order (LO) generator {\sc pythia} 
6.421~\cite{pythia} using MRST 2007 LO*~\cite{lomrst} parton
distribution functions (PDFs).  The production cross-section 
is calculated using {\sc hathor}~\cite{BerlinCS} with approximate next-to-next-to-leading-order (NNLO) QCD calculations
with CTEQ6.6 PDFs~\cite{cteq6} for several heavy-quark masses ($m_Q$). 
In addition, scale uncertainties are evaluated in the range 
$m_{Q}/2$ to $2 \times m_{Q}$ and PDF uncertainties are calculated from the CTEQ6.6 error
eigenvectors.  
The cross-sections and uncertainties for each heavy-quark mass considered in this analysis are shown in Table~\ref{tab:10cross}.
Samples are generated with either $t' \rightarrow Wb$, $t' \rightarrow Ws$, or $t' \rightarrow Wd$ final states;
final results are verified with all three decay modes.

\begin{table}[!htbp]
\centering
\caption{NNLO cross-sections, scale uncertainties and PDF uncertainties
calculated using {\sc hathor} for $Q\bar{Q}$
production in $\sqrt{s} = 7$ TeV $pp$ collisions with CTEQ6.6 PDFs for different heavy-quark mass assumptions.
Other uncertainties related to the cross-sections are neglected.  The 1$\sigma$ uncertainties are each represented by $\Delta$.}
\begin{tabular}{cr@{}lr@{}lr@{}l}
\hline
$m_{Q}$ & \multicolumn{2}{c}{$\sigma_{Q\bar{Q}}$} & \multicolumn{2}{c}{Scale $\Delta$} & \multicolumn{2}{c}{PDF $\Delta$} \\
(GeV) &  \multicolumn{2}{c}{(pb)} & \multicolumn{2}{c}{(pb)} & \multicolumn{2}{c}{(pb)} \\
\hline \\
[-2.4ex]\multirow{2}{*}{300} & \multirow{2}{*}{8} & \multirow{2}{*}{.0} & \hspace{5pt} +0 & .2 & \hspace{5pt} +1 & .0 \\
& \multicolumn{2}{c}{} & $-$0 & .5 & $-$0 & .8 \\ 
[0.6ex]\multirow{2}{*}{350} &  \multirow{2}{*}{3} & \multirow{2}{*}{.2} & +0 & .1 & +0 & .4 \\ 
& \multicolumn{2}{c}{} & $-$0 & .2 & $-$0 & .4 \\
[0.6ex]\multirow{2}{*}{400} &  \multirow{2}{*}{1} & \multirow{2}{*}{.4} & +0 & .1 & +0 & .2 \\  
& \multicolumn{2}{c}{} & $-$0 & .1 & $-$0 & .2 \\
[0.6ex]\multirow{2}{*}{450} &  \multirow{2}{*}{0} & \multirow{2}{*}{.68} & +0 & .02 & +0 & .11 \\ 
& \multicolumn{2}{c}{} & $-$0 & .04 & $-$0 & .09 \\
[0.6ex]\multirow{2}{*}{500} &  \multirow{2}{*}{0} & \multirow{2}{*}{.34} & +0 & .01 & +0 & .06 \\ 
& \multicolumn{2}{c}{} & $-$0 & .02 & $-$0 & .04 \\
\hline
\end{tabular}
\label{tab:10cross}
\end{table}

\subsection{Top quark pair production}

The background due to \ttbar{} production is modeled using
the next-to-leading-order (NLO) generator 
{\sc mc@nlo} v3.41~\cite{MCATNLO} with an assumed top quark mass of 172.5 GeV and the NLO
PDF set CTEQ6.6. The cross-section for \ttbar{} production is
normalized to the value obtained from an approximate NNLO calculation~\cite{BerlinCS}.

\subsection{$Z$ boson, diboson and single-top quark production}

The background from $Z/\gamma^*$ boson production in association with
jets is modeled with the LO generator
{\sc alpgen} v2.13~\cite{AlpgenRef}. 
The LO PDF set CTEQ6.1~\cite{cteq6} is used to generate
$Z/\gamma^*$+jets events 
with dilepton invariant mass \mbox{$m_{\ell\ell} > 10$} GeV.
For $WW$, $WZ$ and $ZZ$ production, events are generated with the LO generator {\sc herwig} v6.5
and the LO PDF set CTEQ6.1.
For the small background from single-top production, {\sc mc@nlo} is used with the NLO PDF set CTEQ6.6, 
invoking the ‘diagram removal scheme’~\cite{drscheme, drscheme2} to remove overlaps between the single-top and \ttbar{} final states. 
The cross-sections for $Z/\gamma^*$+jets samples are determined using NNLO inclusive calculations from {\sc fewz}~\cite{fewz,fewz2}
and from a data-driven technique where possible, 
while the cross-sections for diboson samples are determined using NLO calculations with {\sc mc@nlo}.
The cross-sections for single-top samples are normalized to an approximate NNLO prediction~\cite{singletopCS1,singletopCS2}.

\section{Object Selection}
\label{sec:ObjEvtReco}

Electrons are found by a calorimeter-seeded reconstruction algorithm and must have a track that
matches an energy deposit in the calorimeter.  
They are required to satisfy \mbox{$E_{\textrm{cluster}} / \cosh(\eta_{\textrm{track}}) > 25$ GeV}, where
\mbox{$E_{\textrm{cluster}}$} is the energy deposited in the calorimeter cluster and
\mbox{$\eta_{\textrm{track}}$} is the pseudorapidity of the matching track. 
Electrons are required to be in a pseudorapidity range \mbox{$|\eta_{\textrm{cluster}}| < 2.47$}, excluding 
the transition region \mbox{$1.37 < |\eta_{\textrm{cluster}}| < 1.52$} between the EM calorimeter barrel and endcap.
They must also satisfy a calorimeter isolation
\mbox{$I_{\textrm{cal}} < 3.5$ GeV} requirement in
\mbox{$\Delta R < 0.2$}, where \mbox{$\Delta R = \sqrt{(\Delta\eta)^2 + (\Delta\phi)^2}$}.
Calorimeter isolation is defined as the energy reconstructed within a cone of a certain radius around
the lepton that is not associated with that lepton, and it is represented by $I_{\textrm{cal}}$.
The calorimeter shower shape is required to closely resemble what is expected for electrons~\cite{egamma}.

Jets are reconstructed from topological clusters of energy deposits in the calorimeter~\cite{topocluster} 
using the anti-$k_{\rm t}$ algorithm with distance parameter \mbox{$R = 0.4$}~\cite{antiktalgorithm, antiktalgorithm2}.
These jets are calibrated to the hadronic energy scale using a correction factor obtained from simulation which
depends on \mbox{$p_{\rm T}$} and $\eta$~\cite{EMJES}
They are required to satisfy \mbox{$p_{\rm T} >25$ GeV} and \mbox{$|\eta| < 2.5$}.
Jets that fall within \mbox{$\Delta R < 0.2$} of accepted electrons are rejected.

Muons are found by requiring that a track reconstructed
in the MS has a matching track in the ID.  A loose
cosmic ray rejection is applied by removing all muon pairs that are back-to-back azimuthally (\mbox{$\Delta\phi(\mu,\mu') > 3.1$})
and whose transverse impact parameter with respect to the beam line is greater than 0.5 mm. 
Muon candidates must satisfy \mbox{$p_{\rm T}> 20$ GeV} and \mbox{$|\eta| < 2.5$}.
The muon must be isolated, satisfying calorimeter isolation \mbox{$I_{\textrm{cal}} < 4$ GeV} in \mbox{$\Delta R < 0.3$} and
tracking isolation \mbox{$I_{\textrm{trk}} < 4$ GeV} in \mbox{$\Delta R < 0.3$}.
Tracking isolation is defined as the sum of track momenta within a cone of a certain radius around the lepton vertex,
and it is represented by $I_{\textrm{trk}}$.
The muon must also not fall within \mbox{$\Delta R < 0.4$} of any jet with \mbox{$p_{\rm T} > 20$ GeV}.

The \MET{} is constructed from 
the vector sum of calorimeter topological cluster energies projected onto the transverse plane~\cite{EtMissCite}.
Calorimeter deposits not associated to a jet
are calibrated at the EM energy scale.  
Deposits associated with
selected jets contribute at the corrected hadronic energy scale. 
Muon transverse momenta are included after correcting for muon energy losses in the calorimeters.

\section{Baseline Event Selection}
\label{sec:BaselineEventSel}

The $Q\bar{Q}$ pair decay yields two
charged leptons, two jets and \MET{} from the
undetected neutrinos.  An initial dilepton selection 
is applied to validate
the modeling of the $Z/\gamma^*$ boson production background as well as the
identification of leptons, reconstruction of jets and measurement of \MET{}. 

\begin{figure}[!htbp]
\centering
\subfigure[]{\includegraphics[scale=0.40]{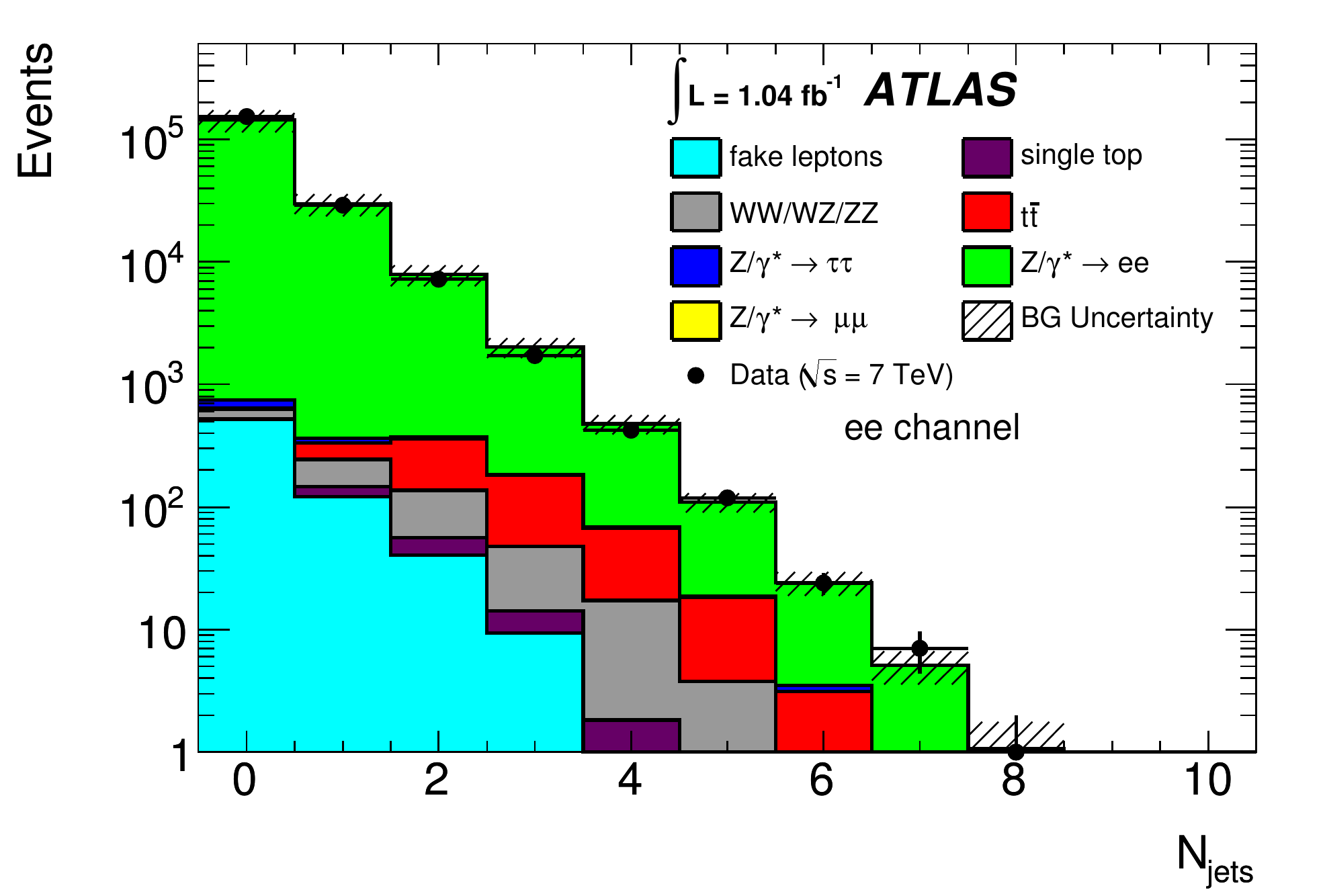}}
\subfigure[]{\includegraphics[scale=0.40]{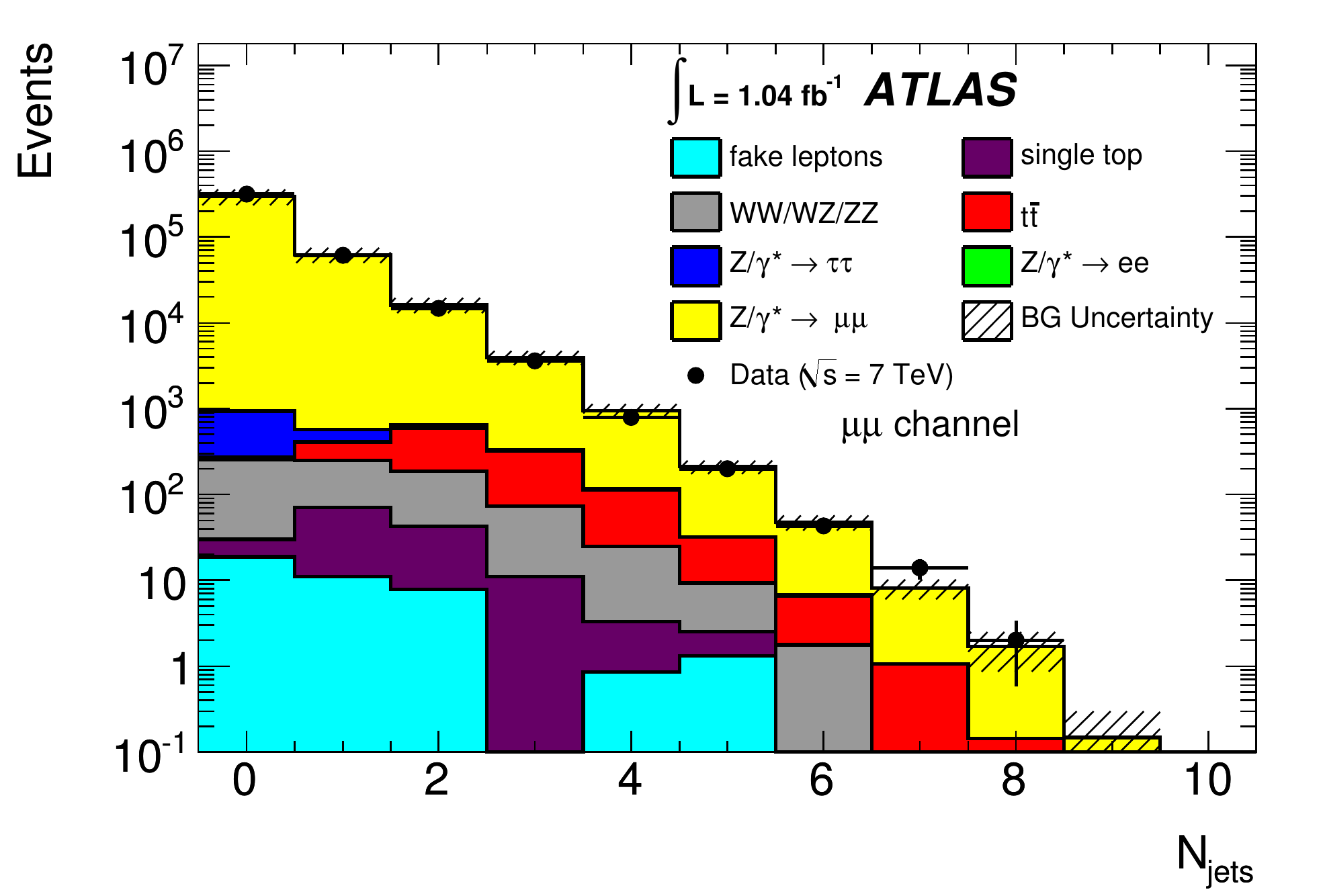}}
\subfigure[]{\includegraphics[scale=0.40]{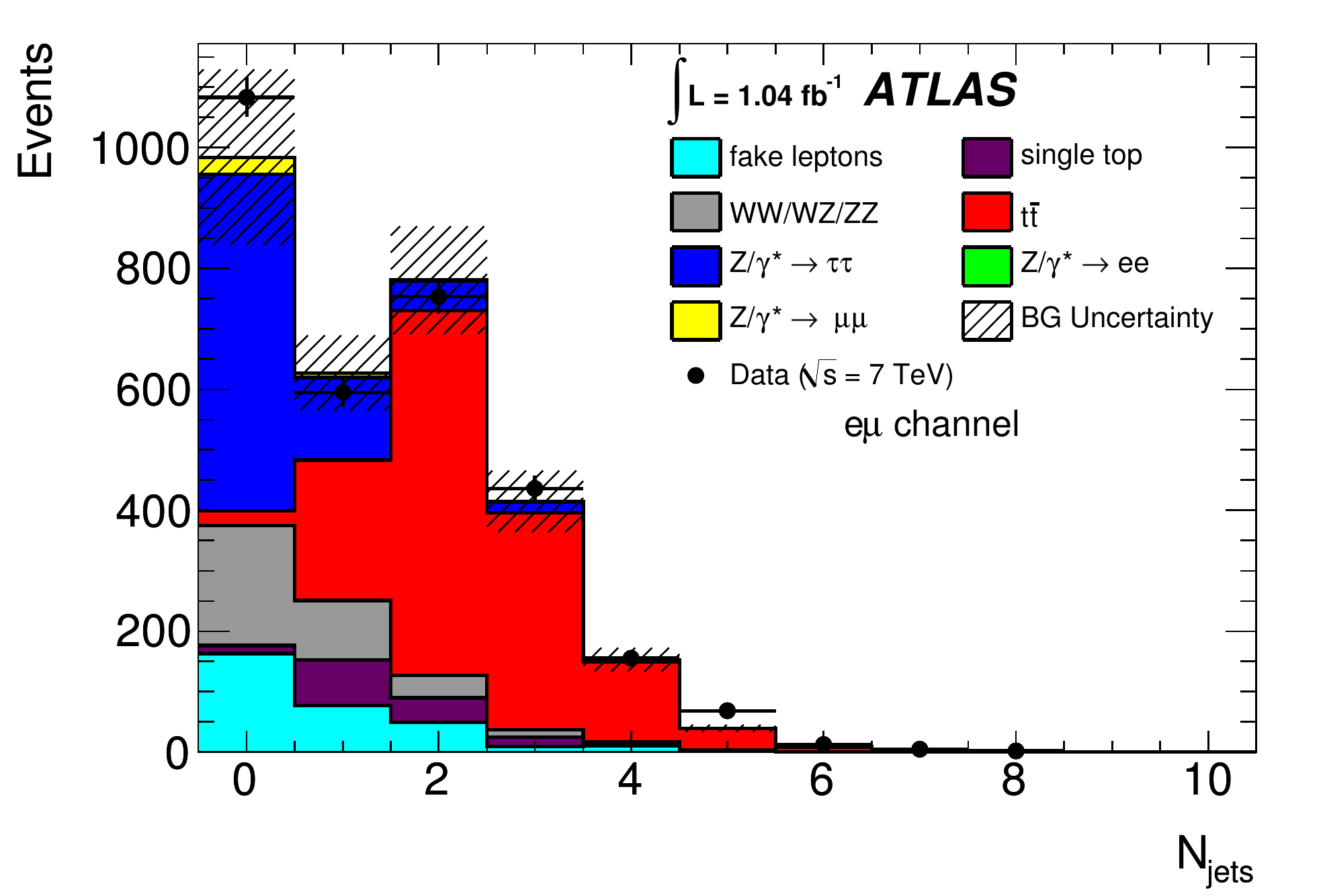}}
\caption{Jet multiplicity after initial dilepton selection in 
(a) $ee$, (b) $\mu\mu$, and (c) $e\mu$ channels.  
  The shaded region indicates the magnitude of cross-section and luminosity uncertainties.
  Samples are stacked in the same order as they are presented in the legend, from left to right; the first entry in the legend is at the bottom of the stack.
  The data-driven Drell-Yan estimate defined in Section~\ref{DDDYSec} has not been applied here.}
\label{fig:dil_njets}
\end{figure}

\begin{figure}[!htbp]
\centering
\subfigure[]{\includegraphics[scale=0.40]{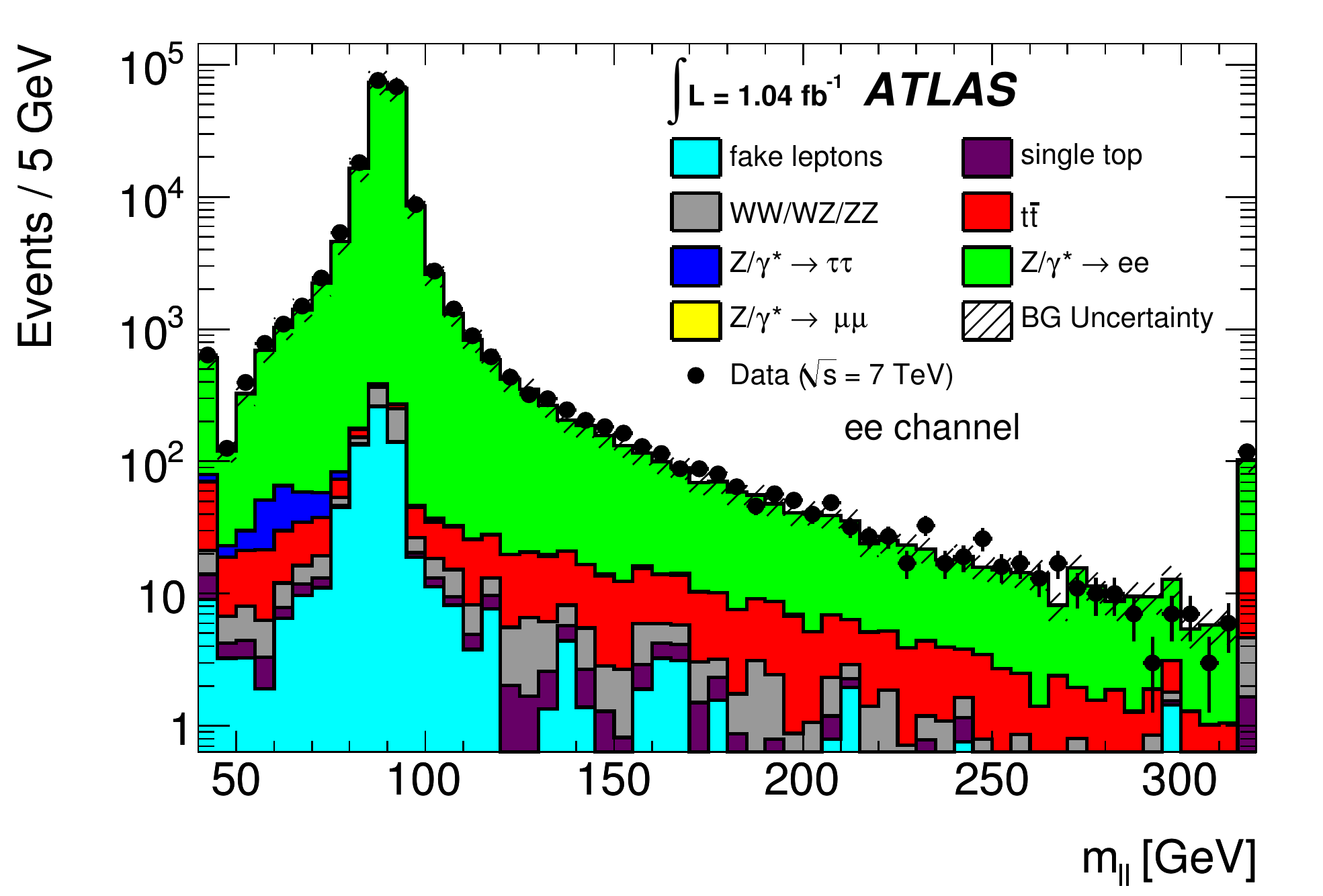}}
\subfigure[]{\includegraphics[scale=0.40]{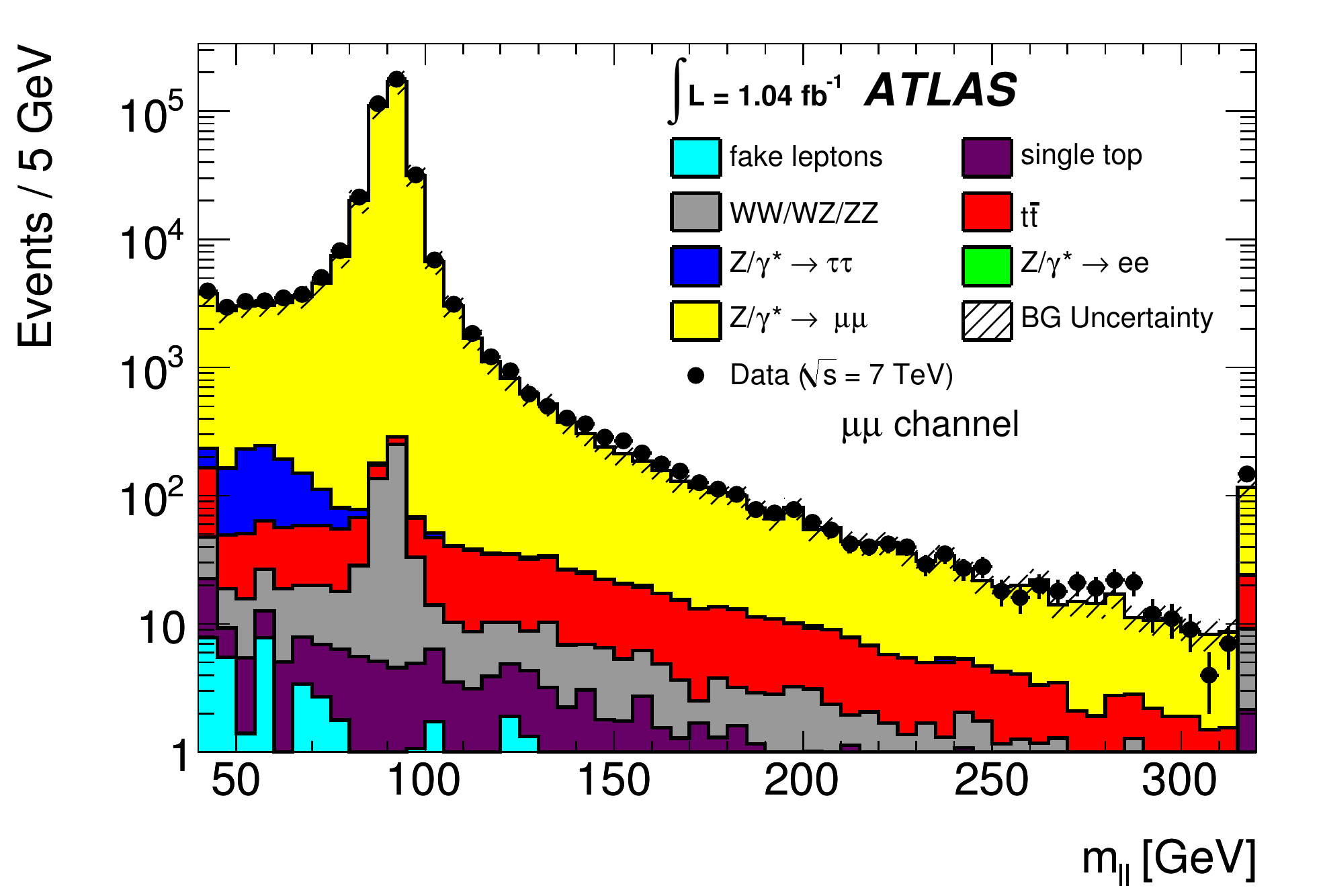}}
\subfigure[]{\includegraphics[scale=0.40]{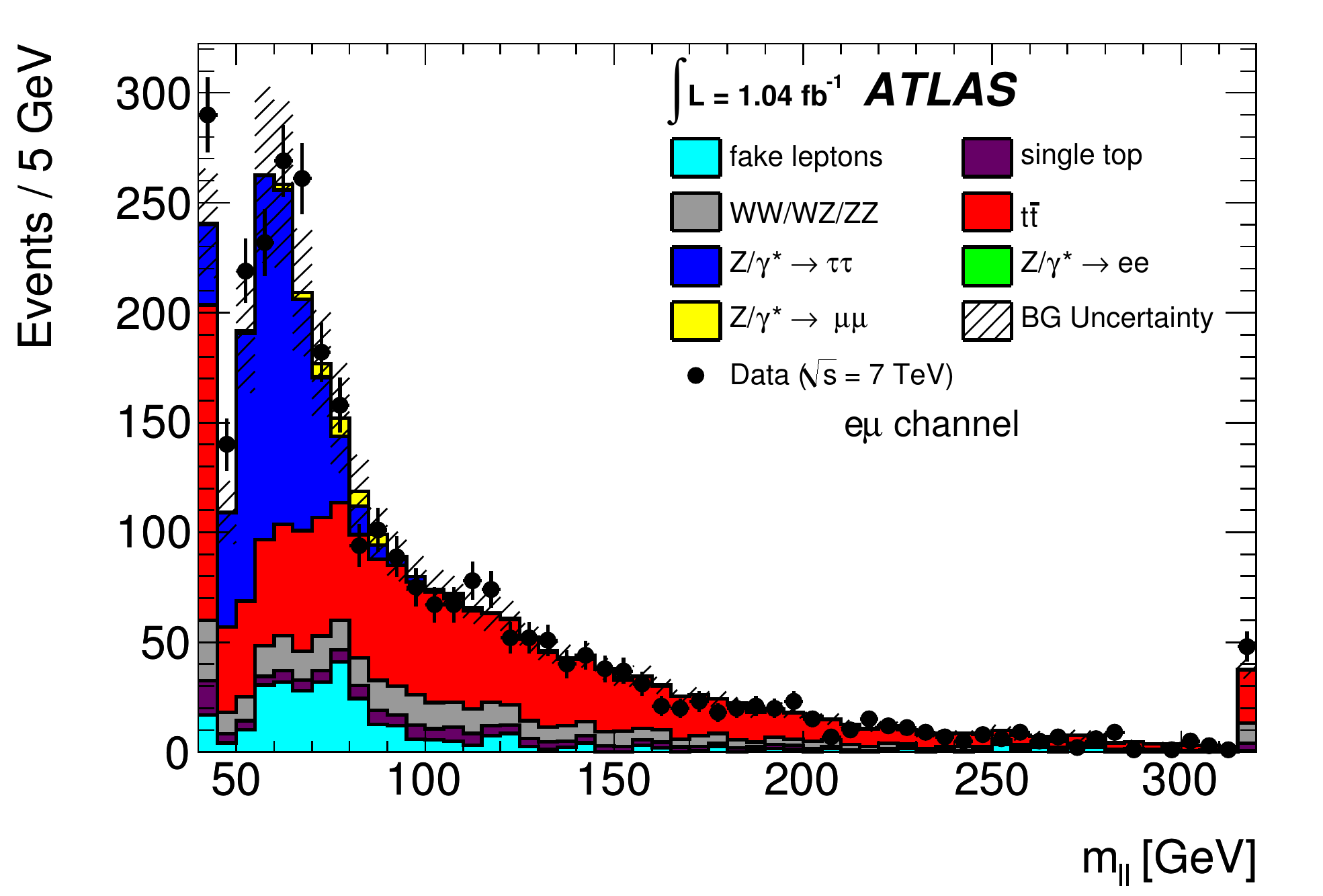}}
\caption{Dilepton mass after initial dilepton selection in
(a) $ee$, (b) $\mu\mu$, and (c) $e\mu$ channels.
  The shaded region indicates the magnitude of cross-section and luminosity uncertainties.
  The last bin contains overflow events, and the first bin contains underflow events.
  Samples are stacked in the same order as they are presented in the legend, from left to right; the first entry in the legend is at the bottom of the stack.
  The data-driven Drell-Yan estimate defined in Section~\ref{DDDYSec} has not been applied here.}
\label{fig:dil_dilepm}
\end{figure}

The initial dilepton selection~\cite{VertexReq,topdilpaper} first requires that an event 
contains a high-quality reconstructed primary vertex.
The event must also have exactly two leptons ($e$ or $\mu$) with opposite charges, 
at least one of which must be associated with the object that triggered the event. 
The two leptons must not share a track in the ID.
At this stage, the data sample is dominated by $Z/\gamma^* \rightarrow \ell\ell$
decays (the Drell-Yan process), although the contribution from \ttbar{} production is evident at large
jet multiplicity, as can be seen in Figures~\ref{fig:dil_njets} and \ref{fig:dil_dilepm}.
These figures show good agreement between the data and background expectation.

To reduce the background from $Z/\gamma^* \rightarrow \ell\ell$ decays, the
baseline selection requires:

\begin{itemize}
\item All events must have at least two jets, each with \mbox{$p_{\rm T}>25$ GeV} and \mbox{$|\eta|<2.5$};
\item Same-flavor events ($ee$ and $\mu\mu$) must satisfy a missing transverse momentum requirement, \mbox{\MET{} $> 60$ GeV};
\item The dilepton invariant mass of same-flavor events ($ee$ and
  $\mu\mu$) must be greater than $15$ GeV and must fall outside a
  window around the $Z$ boson mass, defined as 81 GeV $< m_{\ell\ell} <$ 101 GeV;
\item In different-flavor events ($e\mu$), \HT{}, defined as the scalar sum of $E_{\rm T}$ from every lepton and jet passing the object selection criteria, must exceed $ 130$ GeV.  The \HT{} requirement reduces the $Z/\gamma^* \rightarrow \tau\tau$ background, where a \MET{} requirement is insufficient due to the presence of neutrinos.

\end{itemize}

\section{Data-Driven Estimates}
\label{sec:DDSec}
\subsection{Drell-Yan events}
\label{DDDYSec}

The total number of Drell-Yan $ee$ and $\mu\mu$ events remaining after the baseline selection has been applied
is estimated with a data-driven technique that extrapolates from a control region (CR)~\cite{ttbar690pbCONFnote}.  
Events in the CR have dilepton invariant mass in the range $81$ GeV -- $101$ GeV 
with at least two jets and \mbox{\MET{} $>$ 30 GeV}.
The number of data events in the control region, Data(CR),
and MC $Z/\gamma^*$+jets events in the control region, $\textrm{MC}_{\textrm{DY}}$(CR), 
are used to scale the prediction of $Z/\gamma^*$+jets events in the signal region, $\textrm{MC}_{\textrm{DY}}$.
Non-$Z/\gamma^*$ background processes in the control region, $\textrm{MC}_{\textrm{other}}$(CR), are subtracted from the data using MC predictions.
The estimated number of $Z/\gamma^*$+jets events in the signal region, $N_{\textrm{DY}}$, 
in the $ee$ and $\mu\mu$ channels is calculated with Equation~\ref{eqn:DYDD}:

\begin{equation}
\label{eqn:DYDD}
N_{\textrm{DY}} = \frac{(\textrm{Data(CR)} - \textrm{MC}_{\textrm{other}}(\textrm{CR}))}{\textrm{MC}_{\textrm{DY}}(\textrm{CR})} \times \textrm{MC}_{\textrm{DY}}.
\end{equation}

\subsection{Fake lepton events}
A small fraction of the background consists of events in which a jet or a non-prompt lepton is misidentified as a prompt lepton from $W$ boson decay.
Prompt leptons and misidentified non-prompt leptons are referred to as real and fake leptons, respectively.
Fake muons are predominantly produced from semi-leptonic $b$ or $c$
quark decays in which the muon passes the isolation requirements
despite being produced in association with a jet. 
There are three principal mechanisms for producing fake electrons:
heavy-flavor decay, 
light flavor jets with a leading $\pi^0$ overlapping with a reconstructed
track from a charged particle,
and asymmetric conversion of photons
into $e^+e^-$.  
The largest source of events with fake leptons is $W$ boson
production with associated jets, including lepton plus jets decays of top quark pairs.

A ‘matrix method’~\cite{topdilpaper} is used to estimate the fraction of the sample that
comes from fake lepton events.
A looser lepton selection is
defined, and the number of observed dilepton events with
two tight leptons ($N_{\textrm{TT}}$), one loose and one tight lepton ($N_{\textrm{TL}}$, $N_{\textrm{LT}}$) or
two loose leptons ($N_{\textrm{LL}}$) is counted. 
The leptons are ordered by $p_{\rm T}$ such that the leading lepton 
in $N_{\textrm{TL}}$ is tight and the leading lepton in $N_{\textrm{LT}}$ is loose.
Tight leptons pass the selection criteria defined in Section~\ref{sec:ObjEvtReco}.
Loose electrons need to pass the same selections as the
electrons defined in Section~\ref{sec:ObjEvtReco} except for looser shower shape and calorimeter isolation requirements~\cite{egamma}.
Loose muons only need to satisfy \mbox{$p_{\rm T}> 20$ GeV}, \mbox{$|\eta| < 2.5$} and
the muon-jet overlap requirements defined in Section~\ref{sec:ObjEvtReco}.

The probabilities for real and fake leptons that pass the loose identification criteria 
to also pass the tight criteria are defined as $r_{\ell}$ and $f_{\ell}$, respectively.
These two probabilities are measured separately for \mbox{$\ell = e$} and \mbox{$\ell = \mu$}.
Using $r_{\ell}$ and $f_{\ell}$, linear expressions are  obtained for
the observed yields as a function of the number of events with zero,
one and two real leptons together with two, one and zero fake leptons
($N_{\textrm{FF}}$, $N_{\textrm{RF}}$ and $N_{\textrm{FR}}$, $N_{\textrm{RR}}$; in $N_{\textrm{RF}}$ the real lepton has greater \mbox{$p_{\rm T}$} than the fake lepton, and vice versa for $N_{\textrm{FR}}$):

\begin{equation}
\label{eqn:mm_matrix}
\hspace{-0.5cm}
\begin{bmatrix}
N_{\textrm{TT}} \\ N_{\textrm{TL}} \\ N_{\textrm{LT}} \\ N_{\textrm{LL}}
\end{bmatrix} 
= M \begin{bmatrix}
N_{\textrm{RR}} \\ N_{\textrm{RF}} \\ N_{\textrm{FR}} \\ N_{\textrm{FF}}
\end{bmatrix}
\textrm{,}
\end{equation}

\noindent
where $M$ is a $4\times 4$ matrix containing terms proportional to $r_{\ell}$
and $f_{\ell}$.
The matrix is inverted in order
to extract the real and fake content of the observed dilepton event sample.
The method explicitly accounts for the presence of events with two fake leptons.

The probability ($r_{\ell}$) for a real loose lepton to pass the tight criteria is measured in \mbox{$Z\rightarrow\ell\ell$}
events in data with a tag-and-probe method.  
The probability for a fake loose electron 
to satisfy the tight requirements ($f_{e}$) 
is measured by requiring exactly one loose electron in an event
with \mbox{$\MET{} < 10$ GeV}.
The probability for a fake loose muon
to satisfy the tight requirements ($f_{\mu}$) 
is measured in a control region obtained by requiring exactly one loose muon with
\mbox{$|\Delta\phi(\mu,\MET{})| < 0.5$}. 
The baseline selection requirements from Section~\ref{sec:BaselineEventSel} are not applied when checking these control regions.

\section{Mass Reconstruction}
\label{sec:MassReco}

\begin{figure}[!htpb]
\centering
\includegraphics[scale=0.42]{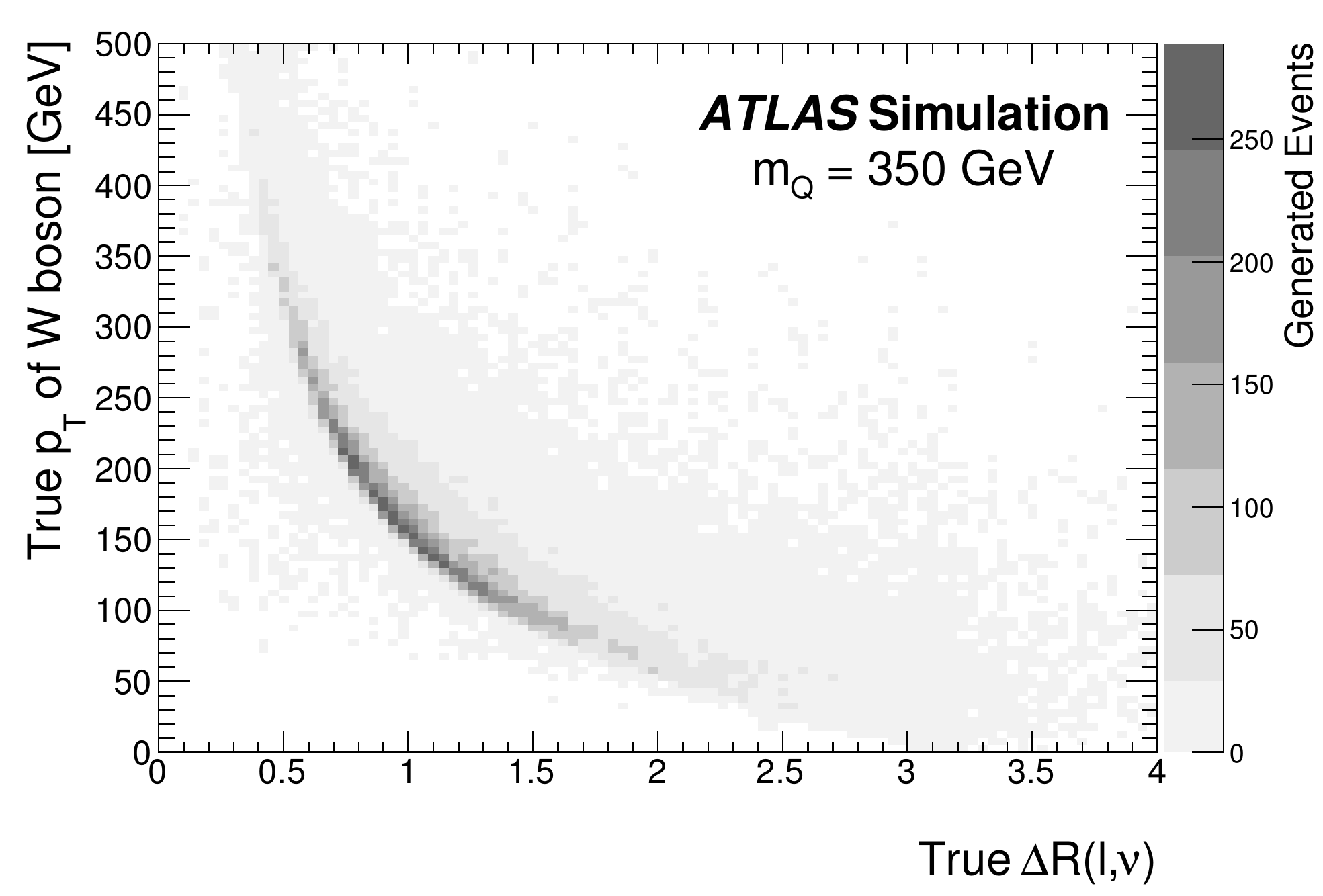}
\caption{True $p_{\rm T}$ of parent $W$ boson versus true $\Delta R$ between its
  daughter lepton and neutrino.
  The scale, shown on the right, indicates the number of generated MC events.}
\label{fig:WdeltaR}
\end{figure}

After the baseline selection has been applied, 
mass reconstruction of heavy quark candidates is performed in order to 
discriminate the heavy-quark decays from the dominant \ttbar{} background.  
Direct reconstruction is not possible, as two neutrinos escape the
detector.  However, a unique feature of the heavy quark
is the large momentum of the daughter $W$ boson,
which makes its decay products approximately collinear in the detector as seen in Figure~\ref{fig:WdeltaR}.

Both neutrino momentum vectors are reconstructed by assuming that the neutrinos are the sole contributors to \MET{}
and that they are approximately collinear with the leptons.
The optimal values of each $|\Delta\eta(\nu,\ell)|$ and each $|\Delta\phi(\nu,\ell)|$ are fit
by minimizing the mass difference between the two reconstructed heavy quarks using \mbox{{\sc minuit}~\cite{MINUITbib}}.
The fitted direction of each neutrino is constrained to be within \mbox{$\Delta R < 2.5$} of the direction taken by
the neutrino's leptonic partner from the $W$ boson decay,
and all jet combinations are considered during each step of the mass difference minimization.
A solution of the minimization procedure is penalized 
if the scalar sum of neutrino momenta exceeds the scalar sum of lepton 
momenta by at least 30\%.  
The square of the difference between each 
reconstructed $W$ boson mass and 80.4 GeV 
is added to the square of the 
heavy-quark mass difference in the minimized function;
the preferred solutions produce $W$ bosons with 
reconstructed masses that are close to the $W$ boson mass.
The full minimization function is: 
$f_{\textrm{min}} = (m_{Q_1} - m_{Q_2})^2 + (m_{W_1} - (80.4 \textrm{ GeV}))^2+ (m_{W_2} - (80.4 \textrm{ GeV}))^2$.



\begin{figure}[!htpb]
\centering
\includegraphics[scale=0.42]{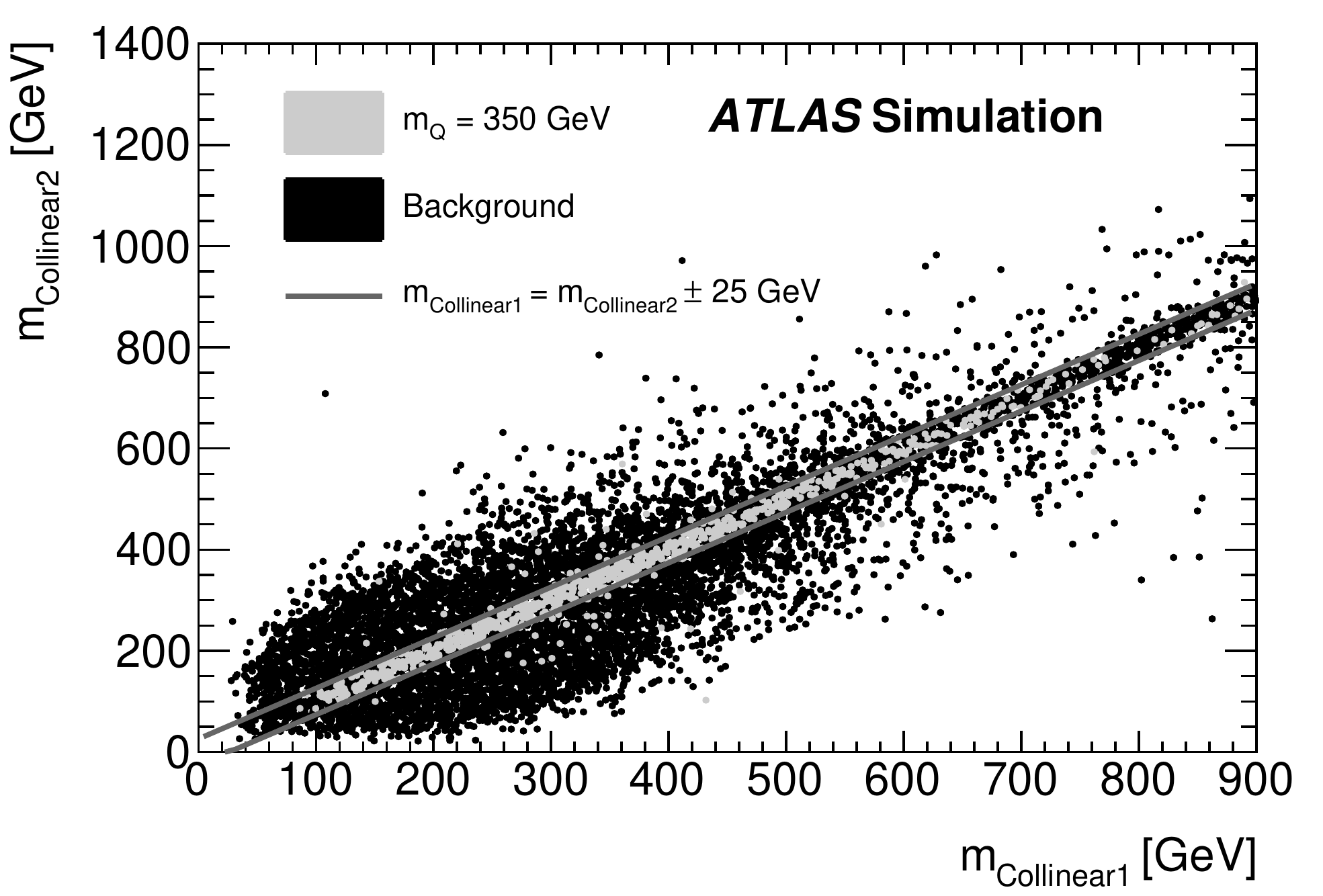}
\caption{
Correlation between reconstructed masses for $Q\bar{Q}$ pairs produced with $m_{Q} = 350$ GeV and for background samples.
The fitting method selects solutions with correlated mass values, however this correlation is smaller for background events.
Events within \mbox{$|m_{\textrm{Collinear}_1} - m_{\textrm{Collinear}_2}| <$ 25 GeV} are kept; 
this region is found between the two lines in the figure.
}
\label{fig:MCollCorrel}
\end{figure}

The two reconstructed mass values tend to be more correlated for signal than background, as shown in Figure~\ref{fig:MCollCorrel}.
This is because the collinear approximation does not work well for single-top, diboson, Drell-Yan, and fake lepton events.
An event is only kept if the two values of reconstructed mass are within 25 GeV of each other.
The selection efficiency for this requirement is greater than 99\% for each signal, 95\% for \ttbar{}, 
and only 75\% -- 90\% for other backgrounds.

\begin{figure}[!htpb]
\centering
\includegraphics[scale=0.40]{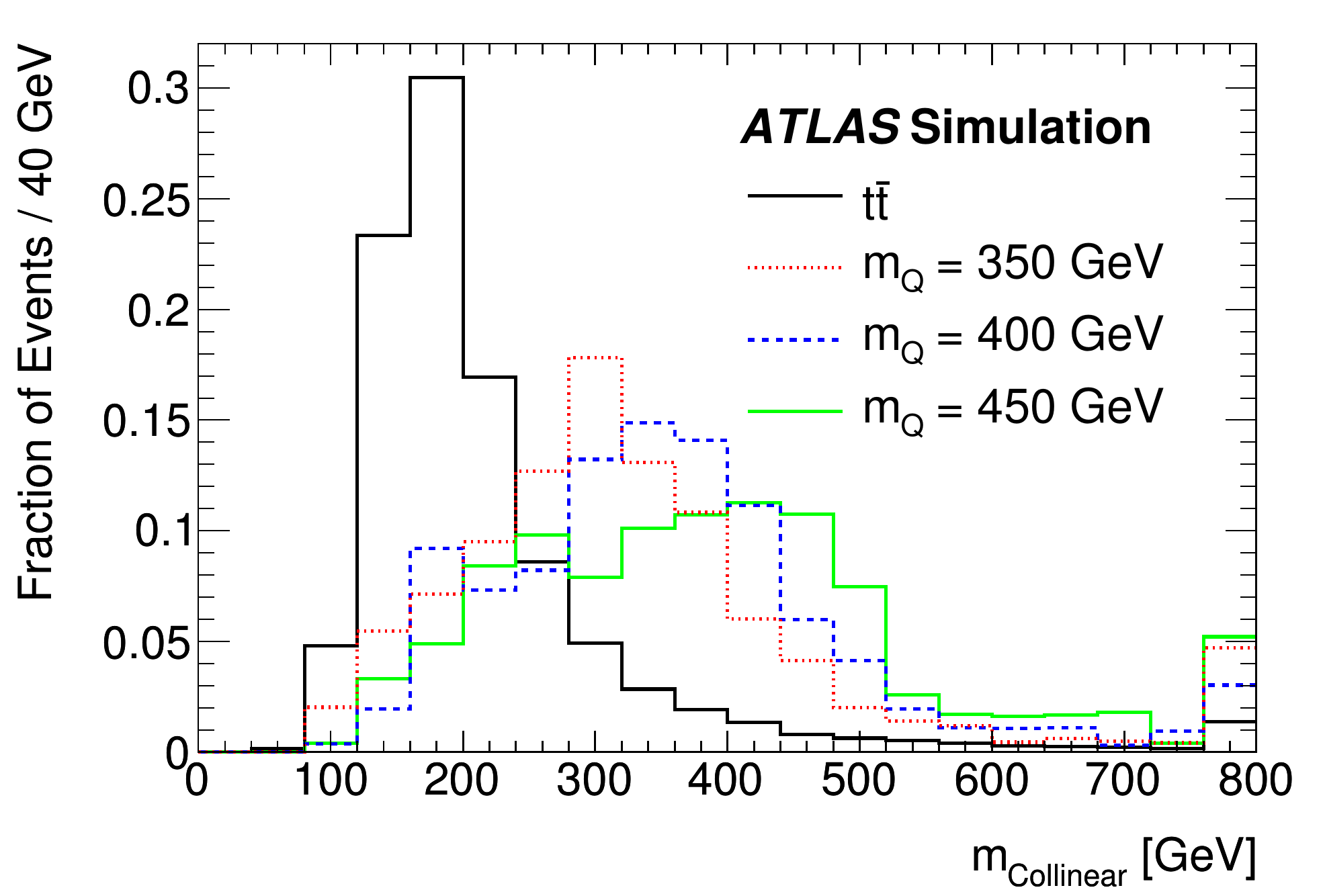}
\caption{$m_{\textrm{Collinear}}$
for simulated
heavy-quark pairs produced with masses of 350 GeV, 400 GeV and 450 GeV and for top quark pairs.
Each histogram is normalized to unit area.  The distributions have long tails that are produced by wrong jet assignment
in events where at least one of the correct jets fails selection requirements.}
\label{fig:shape}
\end{figure}

The final reconstructed mass ($m_{\textrm{Collinear}}$) 
is taken to be the average of the two reconstructed masses in the event.
Distributions of $m_{\textrm{Collinear}}$ for various simulated $Q\bar{Q}$ samples
and the \ttbar{} background are shown in Figure~\ref{fig:shape}.

\begin{figure}[!htbp]
\centering
\subfigure[]{\includegraphics[scale=0.38]{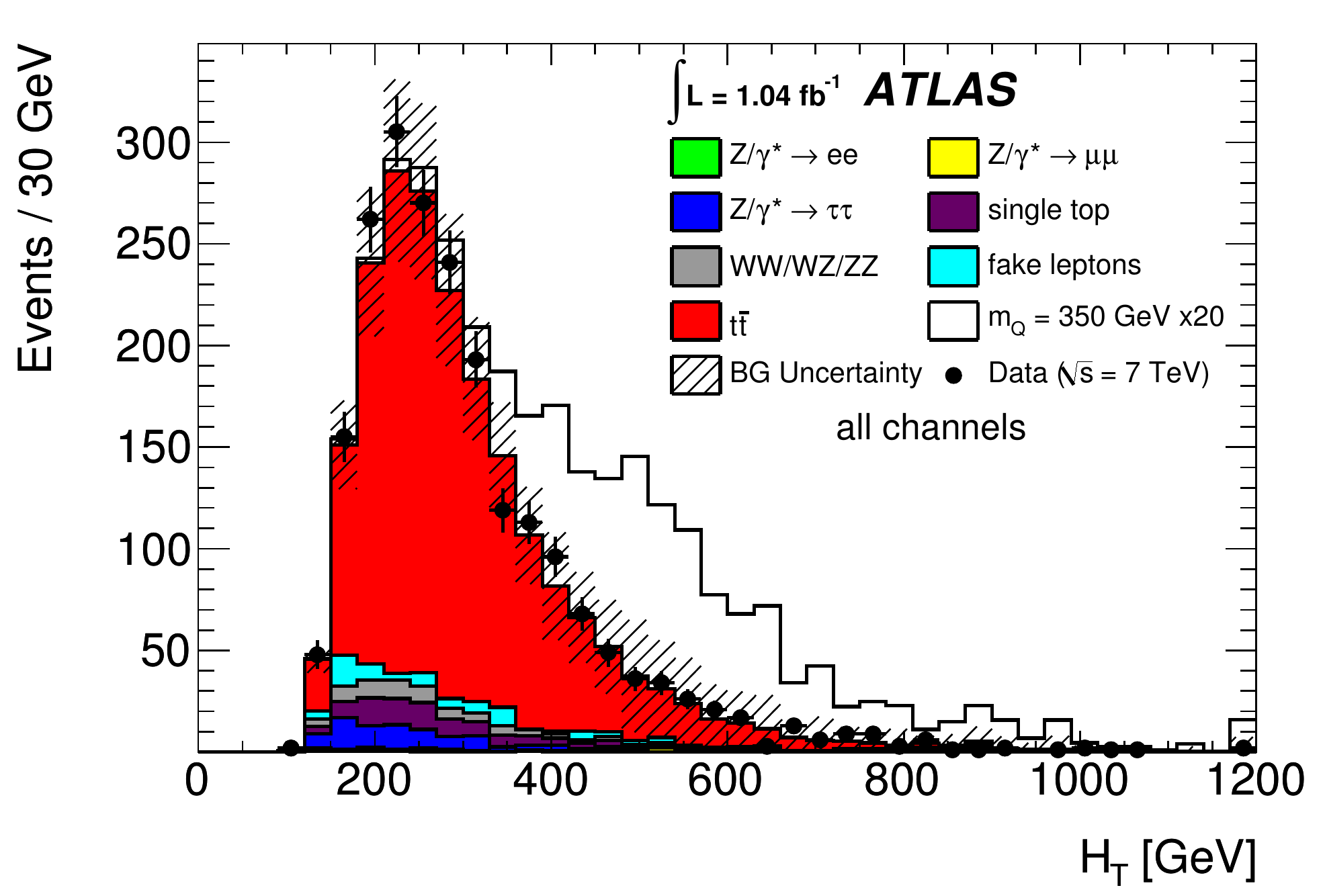}}
\subfigure[]{\includegraphics[scale=0.38]{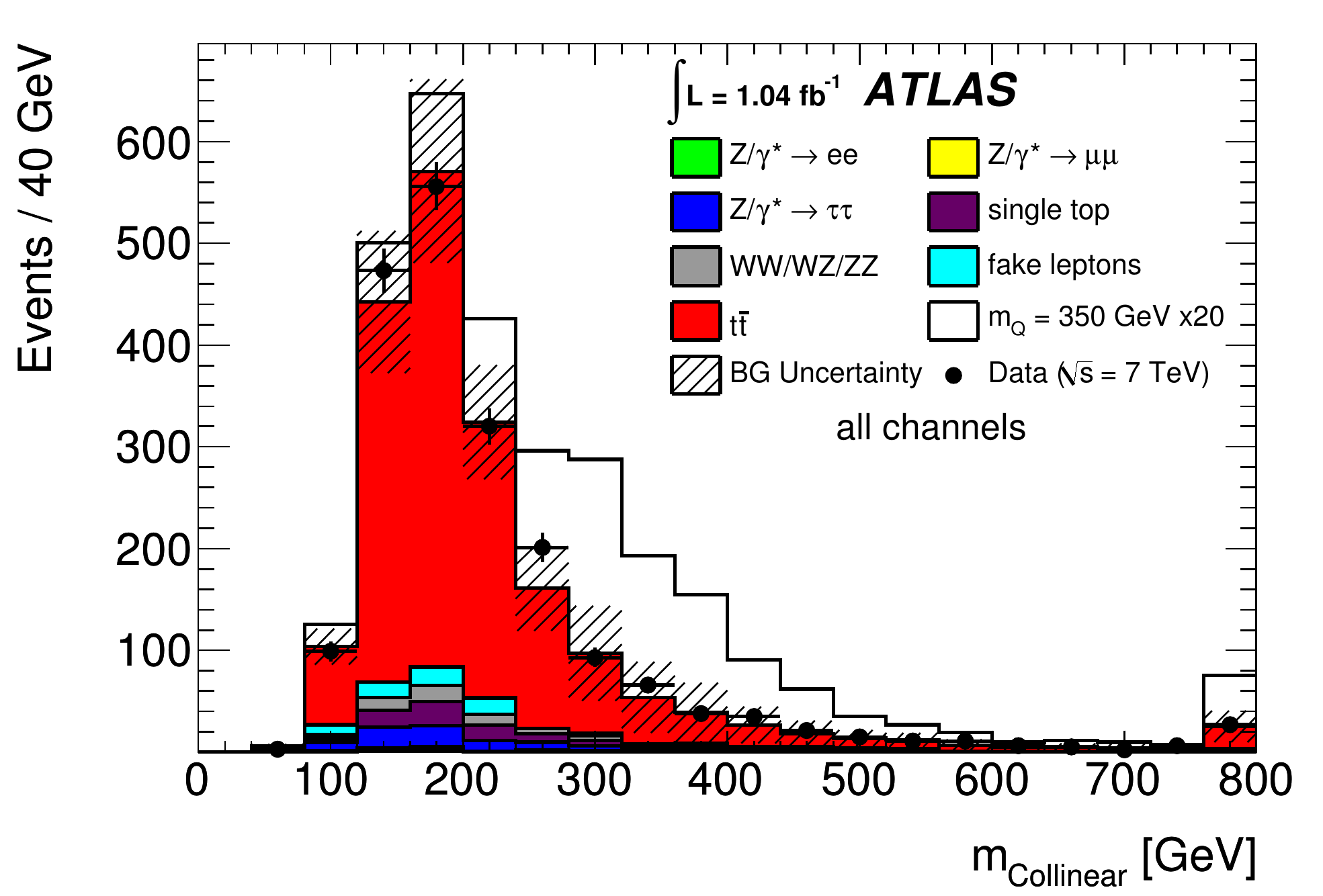}}
\caption{Expected and observed distributions 
of (a) \HT{} and (b) $m_{\textrm{Collinear}}$
for the sum
of $ee$, $\mu\mu$ and $e\mu$ channels after the baseline selection.
The last bin contains overflow events.
Samples are stacked in the same order as they are presented in the legend, from left to right; the first entry in the legend is at the bottom of the stack.
The signal has been amplified to 20 times the expected rate.}
\label{fig:base_ht_mcoll}
\end{figure}

The expected background yields and number of observed events after the
baseline selection are given in Table~\ref{tab:baseline}.
Distributions of \HT{}
and $m_{\textrm{Collinear}}$
are shown in Figure~\ref{fig:base_ht_mcoll}.

\begin{table}[!htbp]
\centering
\caption{Expected and observed number of events after baseline
  selection.  Uncertainties shown are statistical and systematic, added in quadrature.}
\begin{tabular}{lr@{}llr@{}llr@{}ll}
\hline\hline
Process & \multicolumn{3}{c}{$ee$} & \multicolumn{3}{c}{$e\mu$} & \multicolumn{3}{c}{$\mu\mu$} \\
\hline
\\ [-2.4ex] \hline \\
[-2.4ex] \ttbar{} & 190&&$ ^{+40}_{-30}$  & 1140&& $^{+250}_{-200}$  & 370&& $^{+80}_{-70}$ \\
[0.6ex] single-top & 9&.4& $^{+2.2}_{-1.9}$  & 60&& $^{+14}_{-11}$  & 24&& $^{+5}_{-5}$ \\
[0.6ex] $Z/\gamma^*$ $\to$ $ee$ & 6&.3& $^{+2.0}_{-1.9}$  & 0&.0& $^{+0.1}_{-0.0}$  & 0&.0& $^{+0.1}_{-0.0}$ \\
[0.6ex] $Z/\gamma^*$ $\to$ $\mu\mu$  & 0&.0& $^{+0.1}_{-0.0}$  & 2&.2& $^{+1.1}_{-1.1}$  & 17&& $^{+5}_{-4}$ \\
[0.6ex] $Z/\gamma^*$ $\to$ $\tau\tau$ & 7&.3& $^{+2.4}_{-2.2}$  & 62&& $^{+15}_{-12}$  & 16&& $^{+4}_{-4}$ \\
[0.6ex] $WW$, $WZ$, $ZZ$ & 8&.7& $^{+2.2}_{-1.9}$  & 49&& $^{+11}_{-10}$  & 12&.7& $^{+3.0}_{-2.6}$ \\
[0.6ex] fake leptons & 3&.7&  $\pm 2.8$  & 70&&  $\pm 40$  & 0&.5&  $\pm 0.8$ \\
\\ [-2.4ex] \hline \\
[-2.4ex] Total Bg & 230&& $^{+50}_{-40}$  & 1380&& $^{+310}_{-250}$  & 440&& $^{+90}_{-80}$ \\
\\ [-2.4ex] \hline \\
[-2.4ex] Observed & \multicolumn{3}{l}{243} & \multicolumn{3}{l}{1410} & \multicolumn{3}{l}{460} \\
\\ [-2.4ex] \hline\hline
\end{tabular}
\label{tab:baseline}.
\end{table}

\section{Background validation}
\label{sec:BkgVal}

\begin{figure}[!htbp]
\centering
\subfigure[]{%
\includegraphics[scale=0.37]{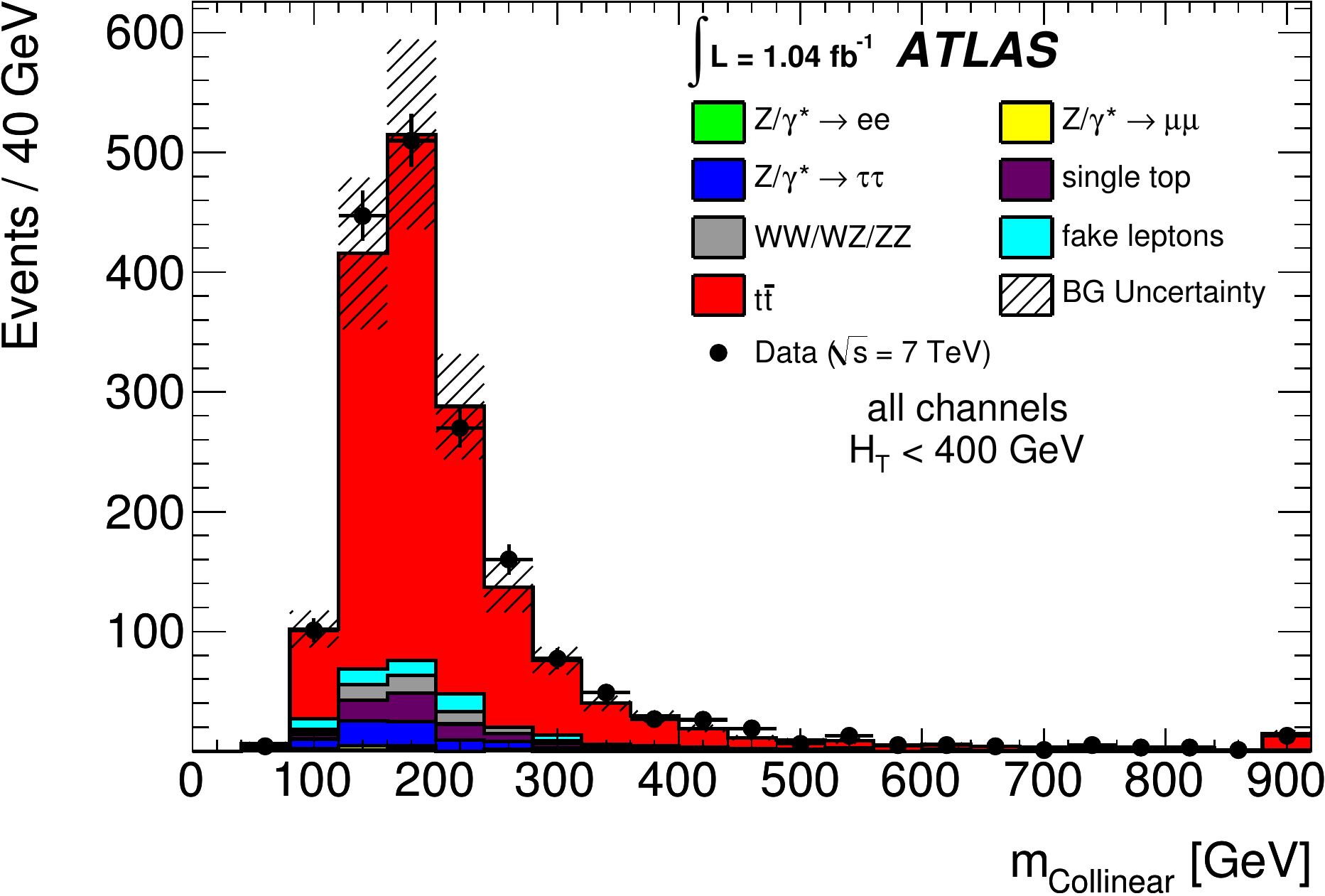}%
}\\
\vspace{-0.2cm}
\subfigure[]{%
\includegraphics[scale=0.37]{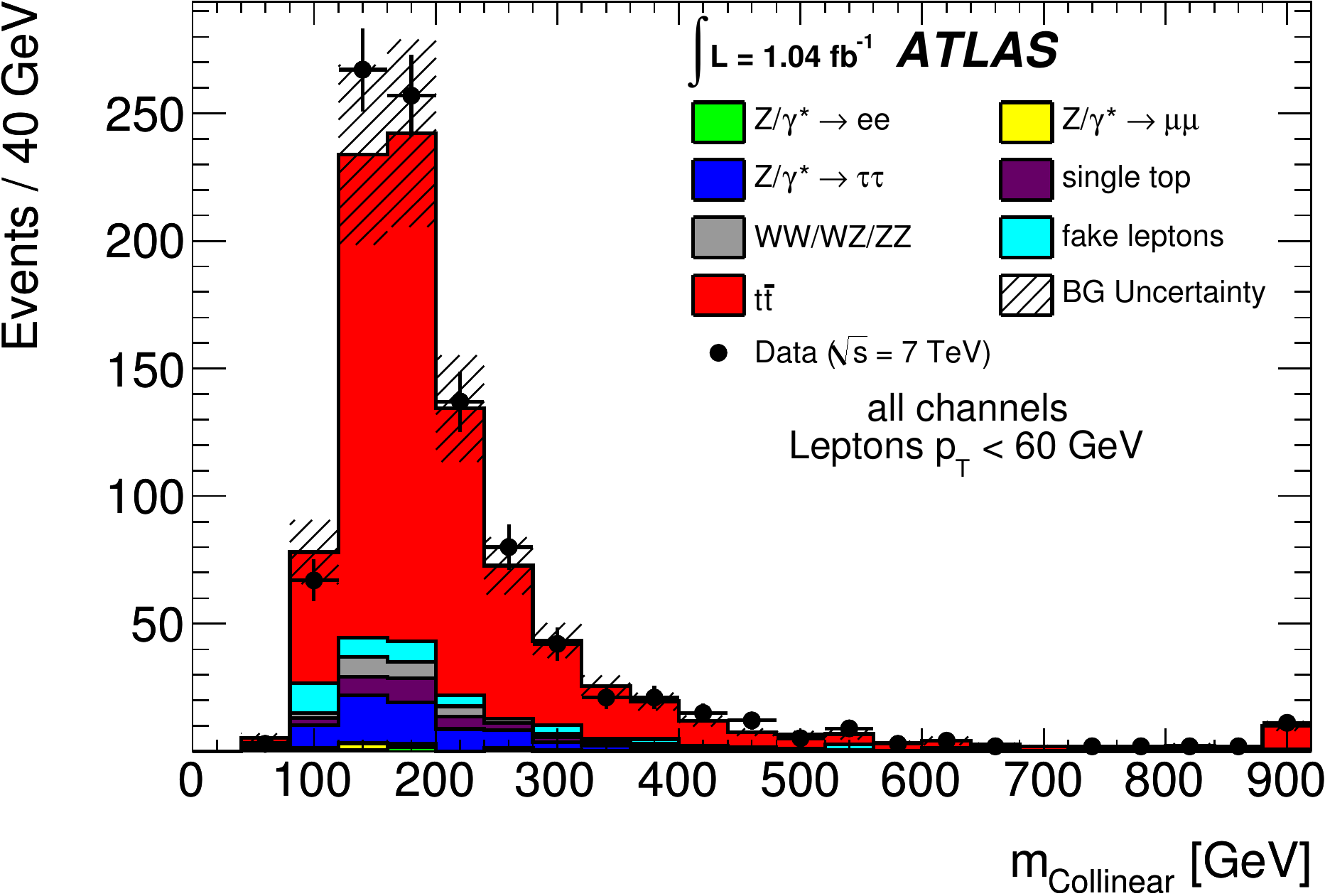}%
}\\
\vspace{-0.2cm}
\subfigure[]{%
\includegraphics[scale=0.37]{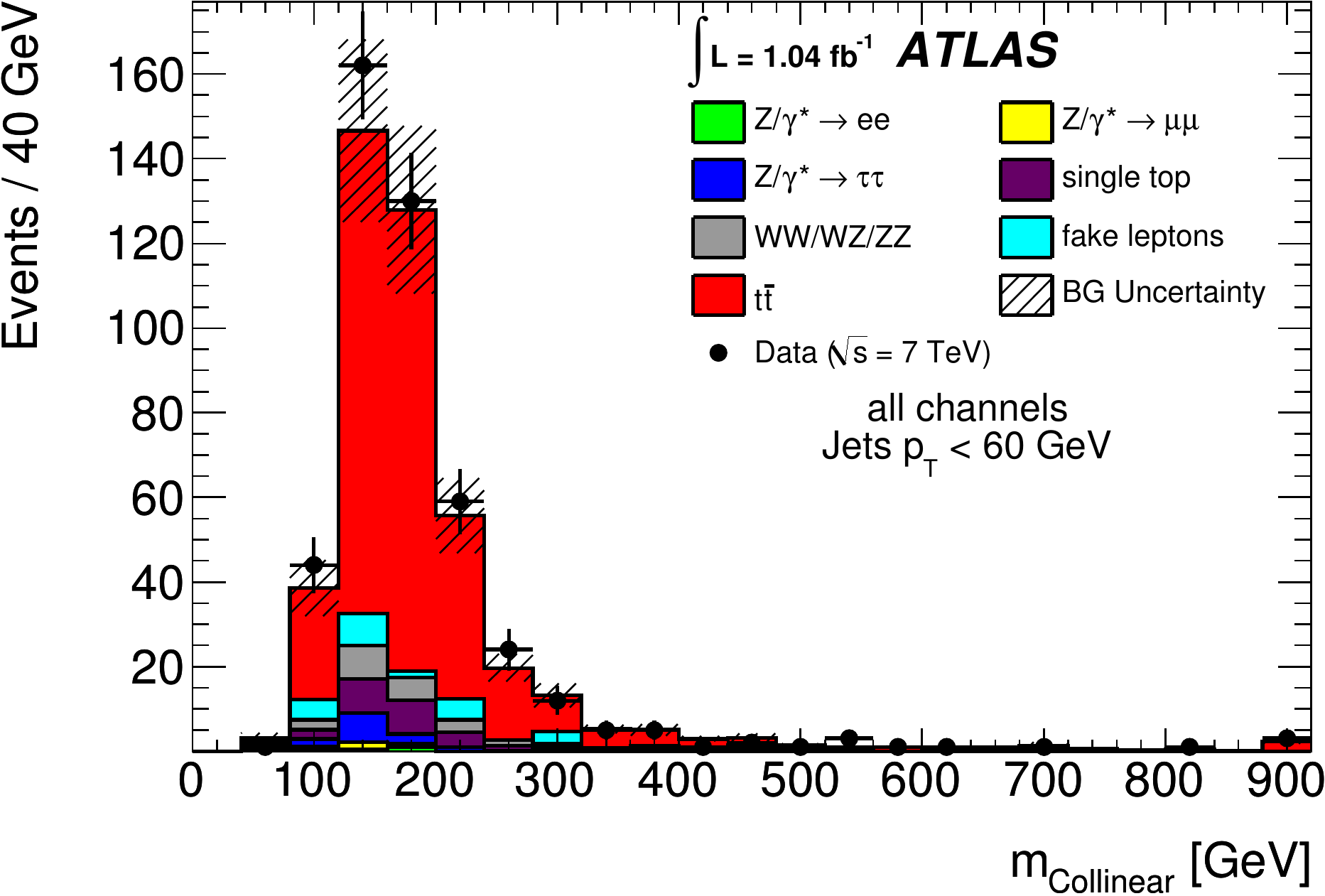}%
}\\
\vspace{-0.2cm}
\subfigure[]{%
\includegraphics[scale=0.37]{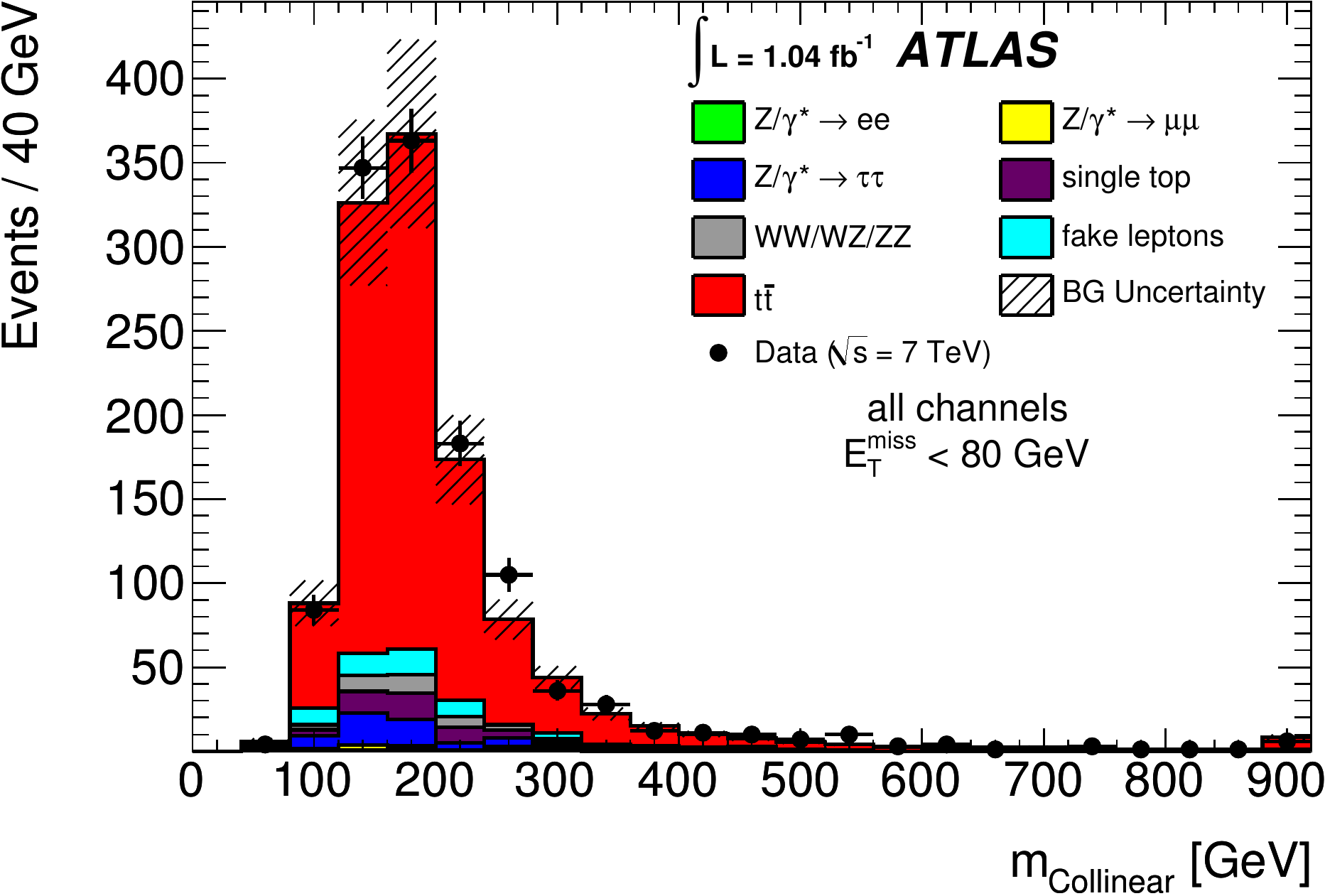}%
}
\vspace{-0.2cm}
\caption{Distributions of $m_{\textrm{Collinear}}$
in events that have 
(a) \mbox{\HT{}$<400$ GeV},
(b) two leptons with \mbox{$p_{\textrm{T}} < 60$ GeV},
(c) all jets with \mbox{$p_{\textrm{T}} < 60$ GeV}
or (d) \mbox{\MET{} $< 80$ GeV}.
Each histogram contains the sum of the $ee$, $\mu\mu$ and $e\mu$ channels.
The last bin contains overflow events.
Samples are stacked in the same order as they are presented in the legend, from left to right; the first entry in the legend is at the bottom of the stack.
}
\label{fig:bg_highht}
\end{figure}

Event samples with the baseline selection and low \HT{}, low lepton $p_{\rm T}$, low jet $p_{\rm T}$ or low \MET{} are examined
to validate the modeling of the background (Figure~\ref{fig:bg_highht}).
These conditions cause the distributions to be depleted of signal.
In each case, the data is described well by the background model, within uncertainties.

\section{Final event selection}
\label{sec:FinalEventSel}

\begin{figure}[!htpb]
\centering
\subfigure[]{\includegraphics[scale=0.44]{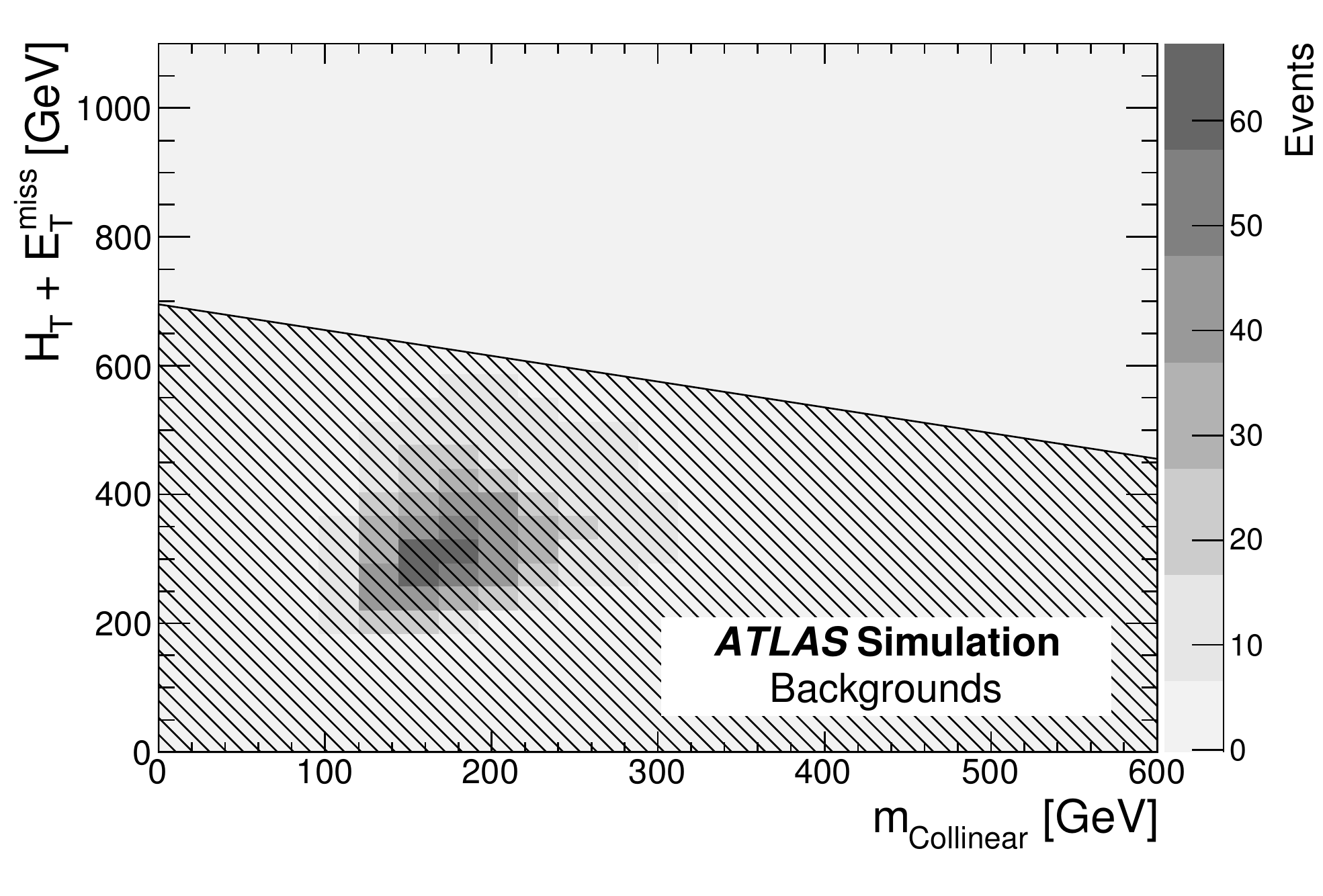}}
\subfigure[]{\includegraphics[scale=0.44]{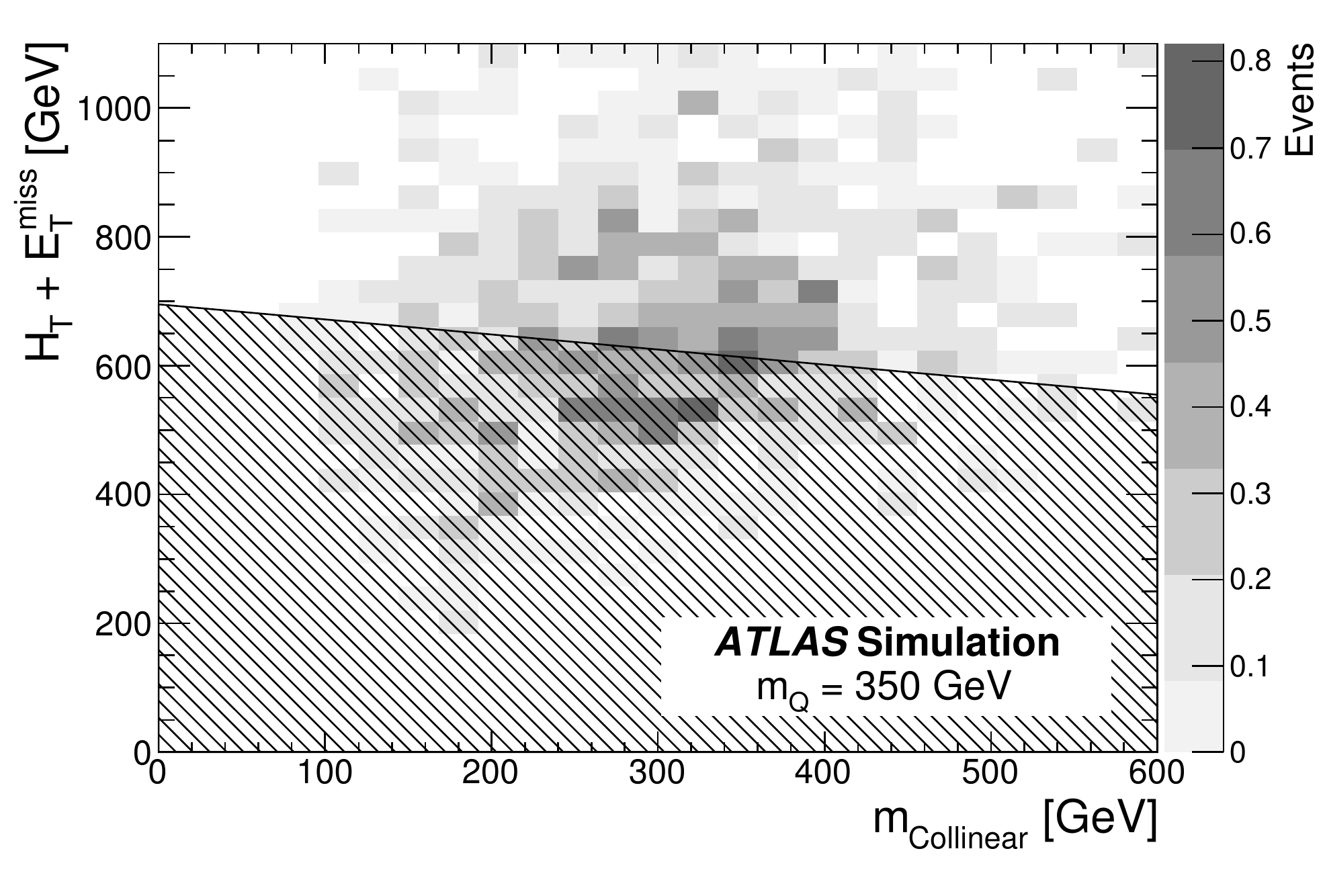}}
\caption{\HT{} + \MET{} versus $m_{\textrm{Collinear}}$ for (a) background and (b) \mbox{$m_Q =$ 350 GeV}
  signal events for the sum of $ee$, $\mu\mu$,
  and $e\mu$ channels.  The scale, shown on the right, indicates the number of reconstructed MC events per bin passing
  the baseline selection and weighted to $\int \mathcal{L}dt =$ \lum{} \invfb{}.
  The shaded region is removed by the triangular selection requirement
  shown in Table~\ref{tab:tricut}.}
\label{fig:HTMColl350}
\end{figure}

\begin{figure}[!htbp]
\centering
\subfigure[]{\includegraphics[scale=0.4]{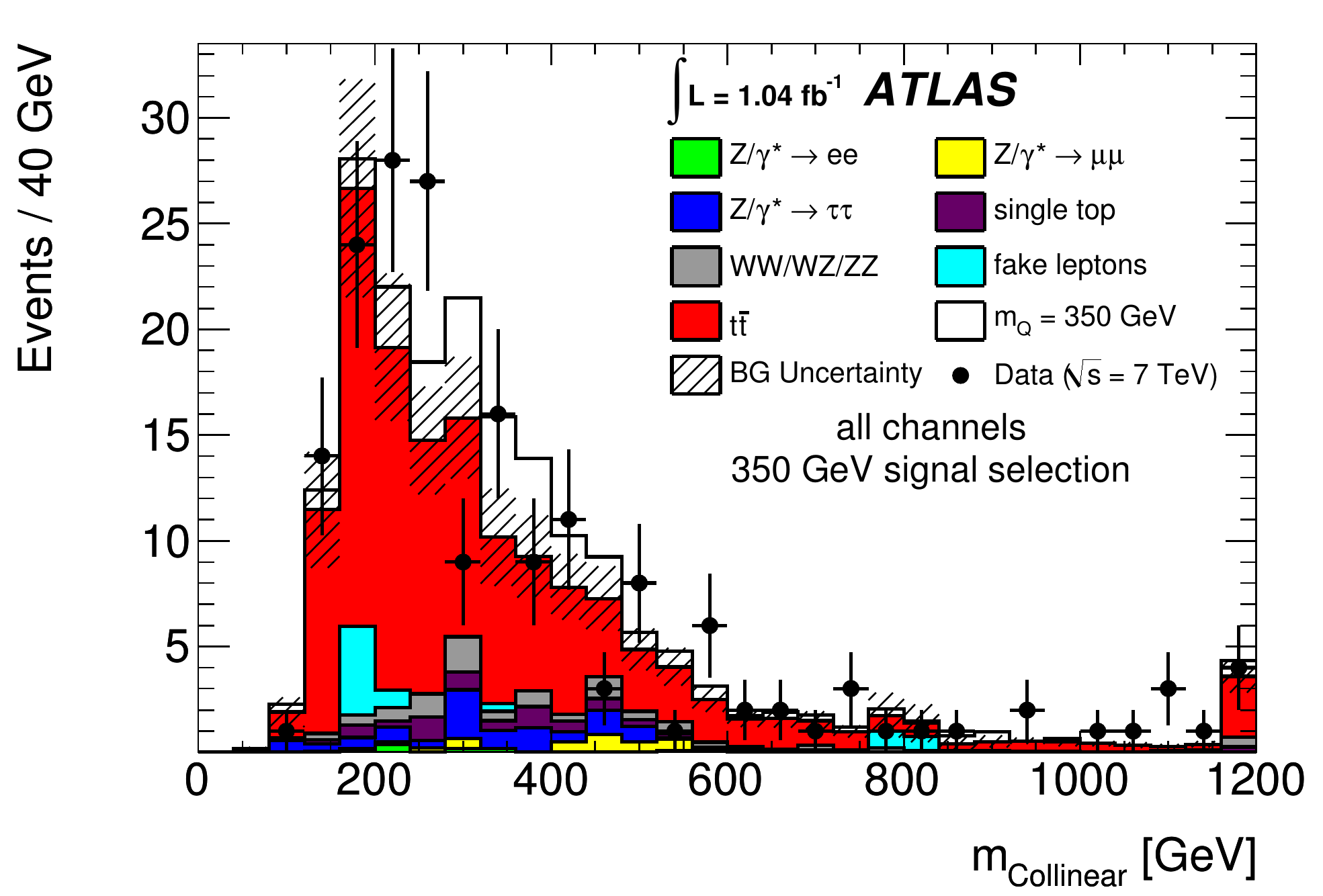}}\\
\subfigure[]{\includegraphics[scale=0.4]{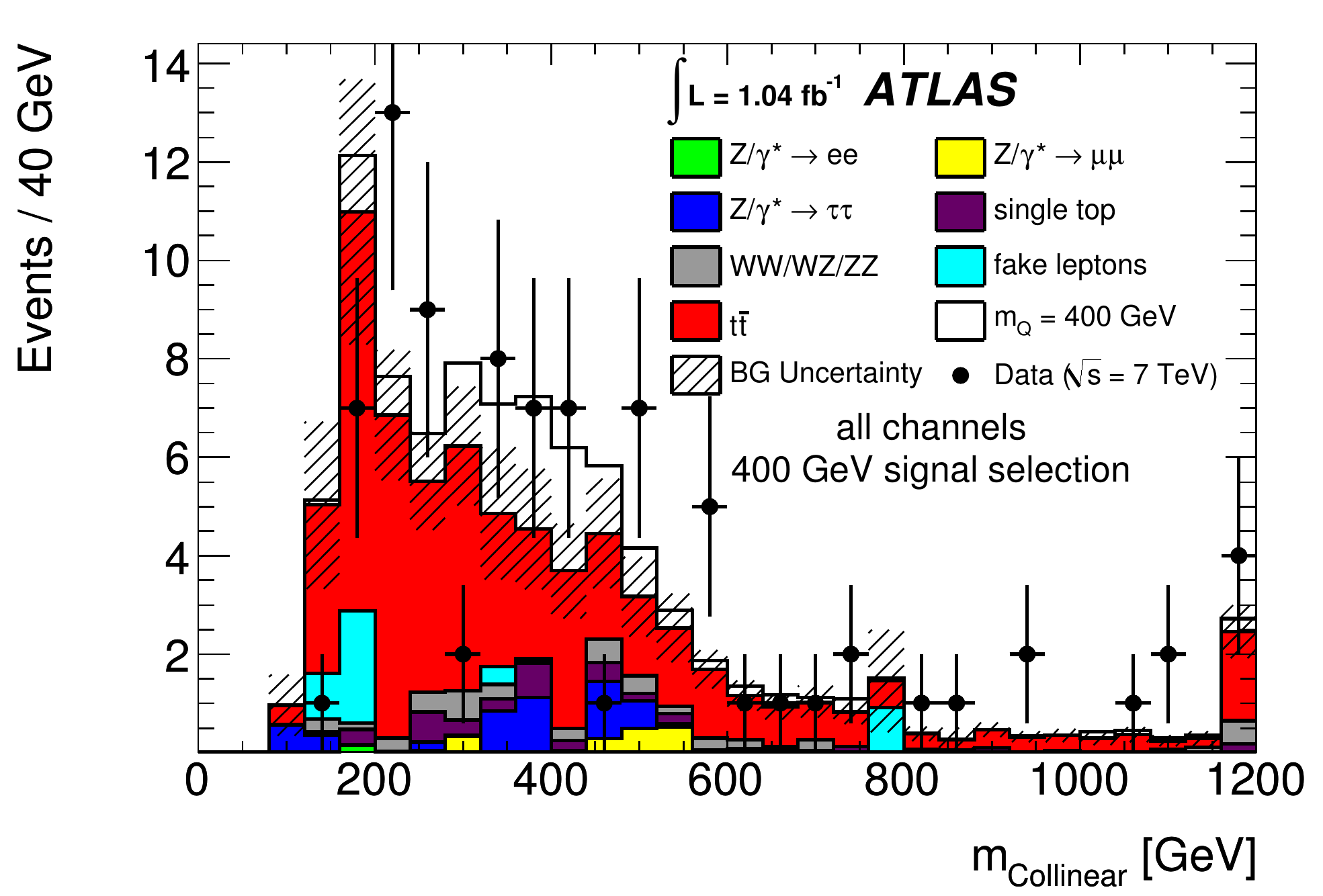}}\\
\caption{Distributions of $m_{\textrm{Collinear}}$ for the sum of $ee$, $\mu\mu$,
  and $e\mu$ channels after applying the final selection for (a) \mbox{$m_Q = 350$ GeV} and (b) \mbox{$m_Q = 400$ GeV}. 
  The last bin contains overflow events.
  The uncertainty bands include all statistical and systematic background uncertainties.  
  The signal samples are normalized using the cross-sections in Table~\ref{tab:10cross}.
  Samples are stacked in the same order as they are presented in the legend, from left to right; the first entry in the legend is at the bottom of the stack.
  }
\label{fig:tricutcompare}
\end{figure}

The baseline selection provides excellent discrimination against
$Z/\gamma^*$ production and other backgrounds, but additional selection requirements are necessary to suppress the dominant \ttbar{} background.
A triangular selection in \HT{} + \MET{} versus $m_{\textrm{Collinear}}$, 
\mbox{\HT{} + \MET{} $> X - 0.4 \times m_{\textrm{Collinear}}$} with $X$ dependent on the assumed signal mass, is applied.
Mass-dependent requirements on \MET{} and leading jet $p_{\rm T}$ are imposed as well.
These selection requirements are optimized in MC simulation by seeking a point of maximum significance, 
$S/\sqrt{S+B}$, while simultaneously varying all of the selection requirement parameters.
Distributions of \HT{} + \MET{} versus $m_{\textrm{Collinear}}$ for background and signal are shown in Figure~\ref{fig:HTMColl350}.
Tables~\ref{tab:tricut} and \ref{tab:tricut2} list the full set of optimized selection requirements at each mass point.
Table~\ref{tab:tricutcompare} lists the
expected backgrounds, expected signal and observed data for each mass point after this final selection.
Figure~\ref{fig:tricutcompare} shows the distributions in $m_{\textrm{Collinear}}$ for two signal samples
after the final selection.

\begin{table}[!htpb]
\centering
\caption{List of \HT{} + \MET{} versus $m_{\textrm{Collinear}}$ requirements for each $m_Q$.} 
\begin{tabular}{l|c}
\hline
$m_Q$ (GeV) & Triangle Requirement (GeV) \\ \hline
300 & \HT{} + \MET{} $ > 610 - 0.4 \times m_{\textrm{Collinear}}$ \\
350 & \HT{} + \MET{} $ > 700 - 0.4 \times m_{\textrm{Collinear}}$ \\
400 & \HT{} + \MET{} $ > 790 - 0.4 \times m_{\textrm{Collinear}}$ \\ 
450 & \HT{} + \MET{} $ > 880 - 0.4 \times m_{\textrm{Collinear}}$ \\ 
500 & \HT{} + \MET{} $ > 970 - 0.4 \times m_{\textrm{Collinear}}$ \\ \hline
\end{tabular}
\label{tab:tricut}
\end{table}

\begin{table}[!htpb]
\centering
\caption{List of jet $p_{\rm T}$ and \MET{} requirements for each $m_Q$.}
\begin{tabular}{l|c|c}
\hline
$m_Q$ (GeV) & Jet $p_{\rm T}$ (GeV) & \MET{} (GeV) \\ \hline
300 & Leading jet $p_{\rm T} >$ ~80 &  --- \\
350 & Leading jet $p_{\rm T} >$ 120 & --- \\
400 & Leading jet $p_{\rm T} >$ 130 & \MET{} $>$ 70 \\
450 & Leading jet $p_{\rm T} >$ 130 & \MET{} $>$ 70 \\
500 & Leading jet $p_{\rm T} >$ 130 & \MET{} $>$ 70 \\ \hline
\end{tabular}
\label{tab:tricut2}
\end{table}

\begin{table}[!htbp]
\begin{center}
\caption{Expected background, expected signal and observed data in $ee$, $\mu\mu$,
  and $e\mu$ channels for
  $m_Q = 300-500$ GeV after final selection. 
  The uncertainties shown include both statistical and systematic contributions.
}
\begin{tabular}{lr@{}llr@{}llc}
\hline\hline
$m_Q$ & \multicolumn{3}{c}{Expected} &\multicolumn{3}{c}{Expected} &Observed \\
(GeV) & \multicolumn{3}{c}{Background} &\multicolumn{3}{c}{Signal} &Data \\ \hline
\\ [-2.4ex] \hline \\
[-2.4ex] 300 & 300&& $^{+40}_{-40}$ & 95&& $^{+14}_{-12}$ & $315$ \\
[0.6ex] 350 & 148&& $^{+22}_{-18}$ & 35&& $^{+5}_{-4}$ & $180$ \\
[0.6ex] 400 & 75&& $^{+11}_{-10}$ & 17&.1& $^{+2.5}_{-2.1}$ & $89$ \\
[0.6ex] 450 & 49&& $^{+8}_{-6}$ & 8&.4& $^{+1.2}_{-1.0}$ & $57$ \\
[0.6ex] 500 & 30&& $^{+5}_{-4}$ & 4&.4& $^{+0.6}_{-0.5}$ & $36$ \\
\\ [-2.4ex] \hline
\hline\hline
\end{tabular}
\label{tab:tricutcompare}
\end{center}
\end{table}

\section{Systematic Uncertainties}
\label{sec:Systematics}

The major sources of systematic uncertainty are due to 
modeling of the signal and most sources of background.

The uncertainties due to simulation of the
lepton trigger, reconstruction and selection efficiencies are assessed
using leptons from $Z \rightarrow ee$ and $Z \rightarrow \mu\mu$ in data~\cite{topdilpaper}.
MC events are corrected for differences in data and simulation.
The statistical and systematic uncertainties in these corrections are included 
in the uncertainties on the acceptance values. 
Uncertainties in the modeling of the lepton 
energy scale and resolution are studied using 
reconstructed $Z$ boson mass distributions.
The jet energy scale (JES) and its uncertainty are derived by
combining information from test-beam data, LHC collision data and
simulation~\cite{EMJES}. The JES uncertainty varies as a 
function of jet $p_{\rm T}$ and $\eta$ and also accounts for the presence of nearby jets and event pileup.
There is additional uncertainty associated with jets originating from $b$ quarks in simulation.
Smaller uncertainties are associated with
the jet energy resolution and jet finding efficiency.

Uncertainties related to the \MET{} arise due to uncertainties associated with low momenta jets, event pileup, and calorimeter energy not associated with reconstructed leptons or jets~\cite{EtMissCite}.
There is also some uncertainty in estimating the effect of a readout problem affecting a subset of the LAr calorimeter channels 
in a part of the data set.

The use of simulated samples to calculate the signal and background
acceptances gives rise to systematic uncertainties from the
generator choice, the amount of initial and final state radiation (ISR/FSR),
and from the PDF choice. The uncertainty due to the choice of
generator for \ttbar{} events is evaluated by comparing the predictions of {\sc mc@nlo}
with those of {\sc powheg}~\cite{powheg} interfaced to either {\sc herwig} 
or {\sc pythia}. 
The uncertainty due to ISR/FSR is evaluated
by studies using the \mbox{{\sc AcerMC}}~\cite{acer} generator interfaced to
{\sc pythia},  varying the parameters controlling ISR and FSR in a
range consistent with experimental
data~\cite{ATLASExpectedPerformance}.
For $Z/\gamma^{*}$+jets events,
the generator uncertainty is evaluated by comparing the predictions of {\sc alpgen} with the PDF set CTEQ6.1
and {\sc sherpa}~\cite{sherpa} with the PDF set CTEQ6.6.
Finally, the uncertainty in the
PDFs used to generate signal, \ttbar{} and single-top events is
evaluated using the procedure
adopted in a measurement of the \ttbar{} cross-section~\cite{topdilpaper}.

The integrated luminosity measurement carries a 3.7\% uncertainty~\cite{LumiPaper, LumiUncertainty}.
Each sample also has an uncertainty
associated with its theoretical cross-section or with its data-driven rate.
For \ttbar{}~\cite{BerlinCS}, 
single-top~\cite{singletopCS1,singletopCS2}
and \mbox{$Z/\gamma^*$ $\to$ $\tau\tau$}~\cite{Zunc}
the rate uncertainty is estimated from theoretical calculations.
\mbox{$Z/\gamma^*$ $\to$ $ee$}, \mbox{$Z/\gamma^*$ $\to$ $\mu\mu$} and fake lepton event rate uncertainties are evaluated with 
the data-driven fitting described in Section~\ref{sec:DDSec}.
The cross-section uncertainty for each signal point comes from {\sc hathor} NNLO calculations and can be found in Table~\ref{tab:10cross}.
The background normalization uncertainties are listed in Table~\ref{tab:CSsyst}.

\begin{table}
\begin{center}
\caption{Overall normalization uncertainties for each background, 
which are either due to cross-section uncertainties or uncertainties related to data-driven methods.}
\begin{tabular}{lr@{}llr@{}ll}
\hline\hline
Background & \multicolumn{3}{r}{$+1\sigma$ Unc.} & \multicolumn{3}{r}{$-1\sigma$ Unc.} \\ \hline \\
[-2.4ex] \ttbar{} & 7 & & \% & 10 & & \% \\
single-top & 7 & & \% & 7 & & \% \\
$Z/\gamma^*$ $\to$ $ee$ & 60 & & \% & 30 & & \% \\
$Z/\gamma^*$ $\to$ $\mu\mu$ & 40 & & \% & 30 & & \% \\
$Z/\gamma^*$ $\to$ $\tau\tau$ & 40 & & \% & 40 & & \% \\
$WW$, $WZ$, $ZZ$ & 5 & & \% & 5 & & \% \\
fake leptons & 50 & & \% & 50 & & \% \\
\hline\hline
\end{tabular}
\label{tab:CSsyst}
\end{center}
\end{table}

The effects of the systematic uncertainties on the overall background yield are summarized in Table~\ref{tab:syst}
for the cuts used for \mbox{$m_Q = 350$ GeV}.

\begin{table}
\begin{center}
\caption{The effects of the $\pm1\sigma$ systematic uncertainties on the overall background yield.}
\begin{tabular}{lr@{}llr@{}ll}
\hline\hline
Source & \multicolumn{3}{r}{$+1\sigma$ Unc.} & \multicolumn{3}{r}{$-1\sigma$ Unc.} \\ \hline
Lepton trigger & 1&&\% & 1&&\% \\
Lepton ID and reconstruction & 2& & \% & $<$ 1& & \%\\
Jet reconstruction & 3& & \% & 3& & \% \\
Jet energy resolution & 2& & \% & 2& & \% \\
Jet energy scale & 13& & \% & 11& & \% \\
$\mu$ momentum resolution & 2& & \% & 2& & \% \\
$\mu$ momentum scale & 1& & \% & 2& & \% \\
$e$ energy resolution & 1& & \% & 1& & \% \\
$e$ energy scale & $<$ 1& & \% & $<$ 1& & \% \\
$e$ isolation pileup term & 1& & \% & 1& & \% \\
$e$ isolation $p_{\rm T}$ term & $<$ 1& & \% & $<$ 1& & \% \\
\MET{} uncertainties & 1& & \% & $<$ 1& & \% \\
LAr readout problem & 2& & \% & 2& & \% \\
ISR/FSR: \ttbar{} & 8& & \% & 5& & \% \\
MC generator: \ttbar{} & 1& & \% & $<$ 1& & \% \\
MC fragmentation/model: \ttbar{} & 1& & \% & 1& & \% \\
Drell-Yan model & 7& & \% & 7& & \% \\
\hline\hline
\end{tabular}
\label{tab:syst}
\end{center}
\end{table}

\section{Results and Discussion}
\label{sec:Results}

A binned maximum-likelihood ratio technique 
is used to fit
distributions of $m_{\textrm{Collinear}}$ to the observed
data in order to measure the most
likely $Q\bar{Q}$ production cross-section, \mbox{$\sigma(pp\rightarrow Q\bar{Q})$}. 
In the fitting technique, all events with \mbox{$m_{\textrm{Collinear}} > 760$ GeV} are considered to belong in the same bin.
A shape fit is performed on background and signal simultaneously; this
allows the background normalization to float, but maintains the
bin-to-bin relationships which defines the background and signal
shapes in $m_{\textrm{Collinear}}$.  This fitting procedure allows for
the possibility of an underestimated or overestimated background in
the signal region. To take into account systematic uncertainties, the signal and background shapes are smoothly deformed in generated samples of pseudo-data by random variations consistent with these uncertainties as shown in Table~\ref{tab:CSsyst} and Table~\ref{tab:syst};
these fluctuations are not constrained by the data.
Most of the systematic uncertainties are assumed to be correlated between signal and background; 
the uncertainties due to cross-section or data-driven estimates, Drell-Yan modeling, 
\ttbar{} modeling, and \ttbar{} generation are uncorrelated between signal and background.

 A small excess is observed over the background expectation for each
 hypothetical $Q$ mass.

Since the statistical interpretation uses the full shape of the signal
and background distributions, the results the results in Table~\ref{tab:tricutcompare} do not
entirely determine the results; only a signal-like excess would weaken
the observed limits. The excess shape in the \mbox{$m_{\textrm{Collinear}}$} distribution is not particularly signal-like and the data are found to be
in better agreement with 
the background-only than the signal+background hypothesis.

\begin{figure}[!htbp]
\centering
\includegraphics[scale=0.4]{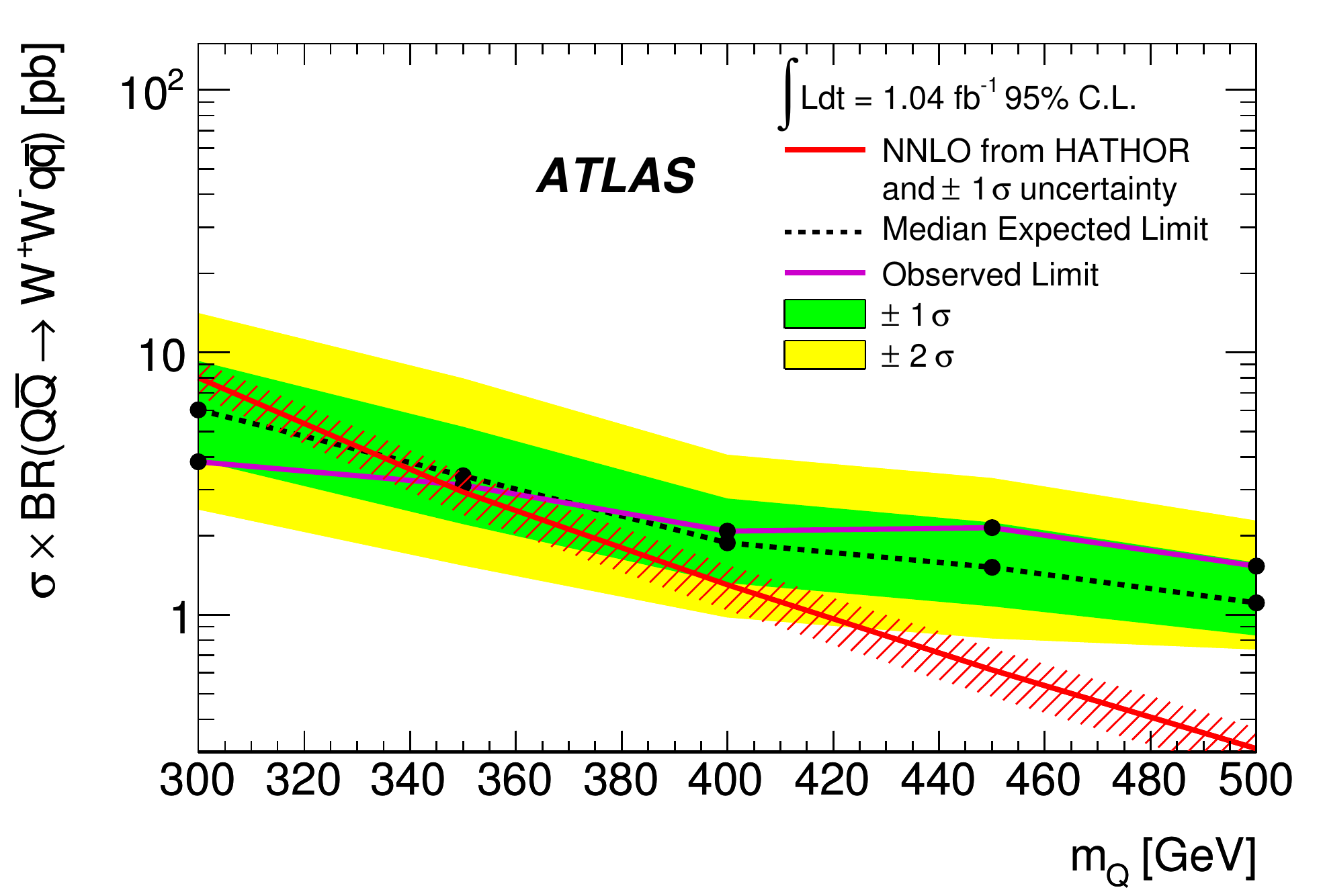}
\caption{ Observed and 
median expected 95\% C.L. cross-section upper limits on $Q\bar{Q}$ production, compared to the theoretical prediction.
The limit was calculated for five signal masses, and a linear interpolation has been made between mass points.
Limits were calculated for simulated samples of $Q \to Wb$, $Q \to Ws$, and $Q \to Wd$,
but the results were approximately the same for all samples.
The results from $Q \to Wu$ and $Q \to Wc$ were assumed to be analogous
for $Q \to Ws$ and $Q \to Wd$, respectively.}
\label{fig:limSyst}
\end{figure}

Statistical interpretation of the fitted cross-section $\sigma$
is made using the CLs technique~\cite{cls1,cls2}.  This technique
performs fits in pseudo-data generated from the background model and
varying levels of injected signal to measure the ability of the fit to
distinguish between the background-only and background-plus-signal
hypotheses. In the case that the data are in better agreement with the
background-only hypothesis, 95\% CL upper limits on the signal
cross-section $\sigma^{95}$ are derived. The limit $\sigma^{95}$ is
chosen so that

\[ \frac{p_{\textrm s}}{1 - p_0} < 0.05 \]

where $p_{\textrm s}$ is the fraction of fits in pseudo-data with
injected signal $\sigma^{95}$ which give a result as seen in the data,
and $p_0$ is the corresponding fraction from pseudo-data drawn from
the background hypothesis.  Thus the performance of the fitting
technique in ensembles of pseudo-data is naturally accounted for.
Figure~\ref{fig:limSyst} shows the observed and expected limits on the
production cross-section \mbox{$\sigma(pp\rightarrow Q\bar{Q})$}.

The upper limit on the production cross-section is converted into a
lower limit on $m_Q$ by finding the point of intersection
with the theoretical prediction as a function of $m_Q$.
This analysis finds a lower limit of \mbox{$m_Q > 350$ GeV} at 95\% confidence level (C.L.)
whereas a limit of \mbox{$m_Q > 335$ GeV} was expected.
This limit assumes that the branching ratio (BR) of $Q \to Wq$ is 100\%.  
Limits were calculated for simulated samples of $Q \to Wb$, $Q \to Ws$, and $Q \to Wd$,
but the results were approximately the same for all samples.  
The results from $Q \to Wu$ and $Q \to Wc$ were assumed to be analogous 
for $Q \to Ws$ and $Q \to Wd$, respectively.

\section{Conclusions}
\label{sec:Conclusion}

This Article presents a search for pair production of heavy quarks decaying to $Wq$ in the dilepton channel at the CERN LHC. 
This search allows 
$q = d$, $s$, $b$ for up-type $Q$ final states or $q = u$, $c$ for down-type $Q$ final states.
The analyzed data correspond to an integrated luminosity of \lum{} \invfb{} collected by the \mbox{ATLAS} detector in $pp$
collisions at \mbox{$\sqrt{s}=7$ TeV}.
To enhance the sensitivity to a new quark,
mass reconstruction is performed by exploiting the boost received by the
heavy-quark decay products.
The reconstructed mass is used for binned maximum-likelihood ratio fitting.

The data are found to be in agreement with the expectation from the Standard
Model. A lower limit is set on the mass \mbox{$m_{Q} > 350$ GeV} at 95\%
confidence level.
This limit assumes BR$(Q \to Wq)=100\%$ and is applicable to many exotic models~\cite{exoquarks,exoquarks2}, 
including up-type fourth-generation quarks $t'$,
down-type fourth generation quarks $b'$, and 
quarks with exotic charges (such as $-4/3$) decaying to light quarks.

~

\section{Acknowledgements}
\label{ack}

We thank CERN for the very successful operation of the LHC, as well as the
support staff from our institutions without whom ATLAS could not be
operated efficiently.

We acknowledge the support of ANPCyT, Argentina; YerPhI, Armenia; ARC,
Australia; BMWF, Austria; ANAS, Azerbaijan; SSTC, Belarus; CNPq and FAPESP,
Brazil; NSERC, NRC and CFI, Canada; CERN; CONICYT, Chile; CAS, MOST and NSFC,
China; COLCIENCIAS, Colombia; MSMT CR, MPO CR and VSC CR, Czech Republic;
DNRF, DNSRC and Lundbeck Foundation, Denmark; EPLANET and ERC, European Union;
IN2P3-CNRS, CEA-DSM/IRFU, France; GNAS, Georgia; BMBF, DFG, HGF, MPG and AvH
Foundation, Germany; GSRT, Greece; ISF, MINERVA, GIF, DIP and Benoziyo Center,
Israel; INFN, Italy; MEXT and JSPS, Japan; CNRST, Morocco; FOM and NWO,
Netherlands; RCN, Norway; MNiSW, Poland; GRICES and FCT, Portugal; MERYS
(MECTS), Romania; MES of Russia and ROSATOM, Russian Federation; JINR; MSTD,
Serbia; MSSR, Slovakia; ARRS and MVZT, Slovenia; DST/NRF, South Africa;
MICINN, Spain; SRC and Wallenberg Foundation, Sweden; SER, SNSF and Cantons of
Bern and Geneva, Switzerland; NSC, Taiwan; TAEK, Turkey; STFC, the Royal
Society and Leverhulme Trust, United Kingdom; DOE and NSF, United States of
America.

The crucial computing support from all WLCG partners is acknowledged
gratefully, in particular from CERN and the ATLAS Tier-1 facilities at
TRIUMF (Canada), NDGF (Denmark, Norway, Sweden), CC-IN2P3 (France),
KIT/GridKA (Germany), INFN-CNAF (Italy), NL-T1 (Netherlands), PIC (Spain),
ASGC (Taiwan), RAL (UK) and BNL (USA) and in the Tier-2 facilities
worldwide.

\onecolumngrid
\clearpage
\input{atlas_authlist} 

\end{document}

%% file: atlas_authlist.tex
\begin{flushleft}
{\Large The ATLAS Collaboration}

\bigskip

G.~Aad$^{\rm 48}$,
B.~Abbott$^{\rm 110}$,
J.~Abdallah$^{\rm 11}$,
A.A.~Abdelalim$^{\rm 49}$,
A.~Abdesselam$^{\rm 117}$,
O.~Abdinov$^{\rm 10}$,
B.~Abi$^{\rm 111}$,
M.~Abolins$^{\rm 87}$,
O.S.~AbouZeid$^{\rm 157}$,
H.~Abramowicz$^{\rm 152}$
% ...
\bigskip

% ...
$^{10}$ Institute of Physics, Azerbaijan Academy of Sciences, Baku, Azerbaijan\\
$^{11}$ Institut de F\'isica d'Altes Energies and Departament de F\'isica de la Universitat Aut\`onoma  de Barcelona and ICREA, Barcelona, Spain\\
% ...
$^{48}$ Fakult\"{a}t f\"{u}r Mathematik und Physik, Albert-Ludwigs-Universit\"{a}t, Freiburg i.Br., Germany\\
$^{49}$ Section de Physique, Universit\'e de Gen\`eve, Geneva, Switzerland\\
% ...
$^{87}$ Department of Physics and Astronomy, Michigan State University, East Lansing MI, United States of America\\
% ...
$^{110}$ Homer L. Dodge Department of Physics and Astronomy, University of Oklahoma, Norman OK, United States of America\\
$^{111}$ Department of Physics, Oklahoma State University, Stillwater OK, United States of America\\
% ...
$^{117}$ Department of Physics, Oxford University, Oxford, United Kingdom\\
% ...
$^{152}$ Raymond and Beverly Sackler School of Physics and Astronomy, Tel Aviv University, Tel Aviv, Israel\\
% ...
$^{157}$ Department of Physics, University of Toronto, Toronto ON, Canada\\
% ...
\end{flushleft}
